\documentclass{revtex4}
\usepackage{amsmath}
\usepackage{amssymb}
\usepackage{graphicx}
\usepackage{color}

\begin{document}

\title{Chameleon stars}

\author{Vladimir Dzhunushaliev,$^{1,2}$
\footnote{
Email: vdzhunus@krsu.edu.kg}
Vladimir Folomeev,$^{2}$
\footnote{Email: vfolomeev@mail.ru}
Douglas Singleton$^{1,3}$
\footnote{Email: dougs@csufresno.edu}
}
\affiliation{$^1$
Institute for Basic Research,
Eurasian National University,
Astana, 010008, Kazakhstan
\\ 
$^2$Institute of Physicotechnical Problems and Material Science of the NAS
of the
Kyrgyz Republic, 265 a, Chui Street, Bishkek, 720071,  Kyrgyz Republic \\
$^{3}$ Physics Department, CSU Fresno, Fresno, CA 93740-8031\\
}

\begin{abstract}
We consider a gravitating spherically symmetric configuration consisting of a scalar field non-minimally coupled to ordinary matter
in the form of a perfect fluid. For this system we find  static, regular, asymptotically flat solutions for both relativistic and
non-relativistic cases. It is shown that the presence of the non-minimal interaction leads to substantial changes
both in the radial matter distribution of the star and in the star's total mass. A simple stability test indicates that,
for the choice of parameters used in the paper, the solutions are unstable.
\end{abstract}

\maketitle

\section{Introduction}

In cosmology scalar fields play a key role in models of both the inflation era of the early Universe \cite{Linde}
and the current accelerated expansion \cite{Copeland:2006wr}. Scalar fields are also widely used to study
smaller scale objects such as the effect of a dark matter, scalar field on the structure of galaxies \cite{Bertone:2004pz}.
Moving to still smaller scales there have been studies of boson stars \cite{Schunck:2003kk} -- compact (usually spherically symmetric)
configurations formed from gravitating scalar field(s).
The sizes of such configurations varies from the microscopic (``gravitational atom'')
up to sizes corresponding to massive black holes such as those presumed to exist at the center of many
galaxies including our own. In this connection, it is quite natural to investigate the role that such scalar fields may have
in the processes of formation and evolution of galaxies and more compact astrophysical structure -- stars and their clusters.
As applied to the stars, it is possible to imagine a situation where scalar fields can exist
inside usual stars consisting of ordinary matter. The existence of such scalar fields would
undoubtedly have an impact on the star's inner structure.
One such example is the model of a star consisting of both ordinary polytropic matter and a ghost scalar field
\cite{Dzhunushaliev:2011xx}.
The presence of the  field leads to the appearance of a tunnel at the center of such stellar configurations
which provides new geometric and physical properties to such objects.
Another example of such stellar objects involves fermion fields interacting with a real scalar field
$\Phi$ not only gravitationally but  also through the Yukawa coupling of the type
$-f \bar{\psi}\psi \Phi$, where $f$ is a Yukawa type coupling constant \cite{Lee:1986tr}.  Further examples
of mixed configurations consisting of both boson and fermion
fields can be found in references~\cite{Henriques:1989ar,Henriques:1989ez,Jetzer:1990xa}.

The investigation in the present work is along the direction of the above works -- we consider a scalar field interacting
non-minimally  with ordinary matter in the form of a perfect polytropic fluid.
An interaction similar to this was used to describe the evolution of dark energy within the framework of chameleon cosmologies
\cite{Farajollahi:2010pk,Cannata:2010qd}.
In chameleon cosmologies the properties of the scalar (chameleon) field depend strongly on the environment
\cite{Khoury:2003rn} into which it is embedded. For example, the mass of the scalar field can change depending on
the background environment \cite{Khoury:2003rn}. As will be shown below, the presence of such
an interaction between the polytropic fluid and the scalar field results in a substantial change of the inner structure of
the polytropic star. This change is related to the fact that the behavior of the scalar
field depends  the ``environment'' (i.e. the fluid) into which it is embedded as is the case with chameleon cosmology models.
Thus we call these configurations {\it chameleon stars}.

In building our model of a chameleon star we will use a real scalar field. In the bulk of the
literature complex (charged) scalar fields are more often used in constructing models of boson stars.
The charge of the complex scalar field generically leads to repulsion which counteracts the attraction of
gravity and thus gives a physical basis for the existence of these boson star solutions.
Real scalar fields have received less attention because they do not carry a charge,
and constructing regular stable solutions is more difficult task in this case.
The known static solutions for real scalar fields are either singular \cite{Wyman:1981bd, jetzer},
or use phantom scalar fields \cite{Kodama:1978dw, Kodama:1979, Dzhunushaliev:2008bq}. Singular solutions with
trivial topology and non-phantom fields have been found for both massless fields \cite{Wyman:1981bd} and for self-interacting fields
\cite{jetzer}. The latter ordinary scalar field solutions are asymptotically flat but the issue of their stability is not quite clear
(for further discussion see reference~\cite{clayton}).
In \cite{Torii:1999uv} regular, static solutions
in the presence of the cosmological constant were found. In this case the space-time asymptotically approaches
de Sitter space-time. The linear stability analysis of these solutions indicate that they are unstable. In contrast
the paper \cite{Dzhunushaliev:2010bv} found regular, static solutions which asymptotically approached anti-de-Sitter space-time, which
were stable under a linear stability analysis.
In the present paper the real scalar field non-minimally interacts with the polytropic fluid. As will be shown below,
this leads to the existence of regular, asymptotically flat solutions.

In the {\it absence} of gravity, Derrick's theorem \cite{derrick,rajaraman} forbids the
existence of non-trivial, stable, regular static $D\geq 3$ dimensional solutions from only scalar fields.
This theorem assumes that the potential energy  of the system is some non-negative function, $V(\varphi_i)>0$,
vanishing only at its absolute  minima. A partial extension of Derrick's theorem to the case of general relativity has been studied
in \cite{Bronnikov:2002kw,Bronnikov:2005gm} where it was shown that the existence of asymptotically flat
particle-like solutions with a regular center and normal (non-phantom) scalar fields is also impossible if
$V(\varphi)>0$. The distinction between the non-gravitational form of Derrick's theorem \cite{derrick,rajaraman}
and the gravitational variant \cite{Bronnikov:2002kw,Bronnikov:2005gm} is that the latter does not touch on the
stability of the solutions.

If one allows other fields besides just scalar fields (e.g. vector or fermion fields) then there are static solutions
when the spatial dimensionality is three or more. In the absence of gravity, one has the 't Hooft-Polyakov monopole which is
a finite energy, stable solution for Yang-Mills vector fields plus scalar fields \cite{tHooft,Polyakov}.
In the presence of gravity plus a real scalar field there are boson star solutions (for a review see \cite{Schunck:2003kk})
which however have one or more of the bad features listed above (e.g. non-trivial topology, not asymptotically flat,
not stable)

In this paper we study the gravitating system of a real scalar field plus a perfect fluid and show that for this
system there are regular, three-dimensional, static solutions.  The most important feature of this system,
aside from the gravitational interaction, is the presence of the non-minimal coupling  between the scalar field
and the perfect fluid. It is this interaction which allows us to avoid the restrictions of the gravitational version
of Derrick's theorem \cite{Bronnikov:2002kw,Bronnikov:2005gm},
and find regular static solutions even if the potential energy of the scalar field satisfies $V(\varphi)>0$.
We take a polytropic equation of state for the perfect fluid. Stellar models using polytropic fluids were investigated
in detail both within the framework of Newtonian gravitational theory \cite{Chandra:1939} and
for strong gravitational fields \cite{Tooper:1964}. Both of these studies hinted at the existence of finite size, regular
solutions. These solutions were successfully used to give a description of both non-relativistic stars and stars
where relativistic effects were important e.g. neutron stars. As will be shown below, the inclusion of the non-minimal
interaction between the polytropic fluid and the scalar field leads to regular solutions which, however, are generally
different from usual polytropic stars.

The paper is organized as follows: In section \ref{gen_equations_cham_star} the general equations
describing a static configuration consisting of a real scalar field coupled to a perfect polytropic fluid are derived.
In section \ref{static_sol_cham_star} these
equations are written for a particular case when the potential energy of the scalar field is chosen
to have a quadratic mass term and a quartic self-interaction term. For this potential energy,
in section \ref{num_calc} we give the results of the numerical calculations for this potential for both
relativistic and non-relativistic cases, and discuss the issue of the stability of the solutions obtained.
Next, in section \ref{non_rel_analyt} we present a simple analytical solution
for the non-relativistic case with the scalar field taken to be massless and for a special choice of the
coupling function. Finally, in section \ref{sec_concl} we summarize the main results and give some
speculations about the physical applications of these chameleon star configurations to neutron stars and
living stars (i.e. stars still on the main sequence).

\section{Derivation of equations for a static configuration}
\label{gen_equations_cham_star}

As discussed in the introduction we consider a gravitating system of a real scalar field coupled to a perfect fluid.
The Lagrangian for this system is
\begin{equation}
\label{lagran_cham_star}
L=-\frac{R}{16\pi G}+\frac{1}{2}\partial_{\mu}\varphi\partial^{\mu}\varphi -V(\varphi)+f(\varphi) L_m ~.
\end{equation}
Here $\varphi$ is the real scalar field with the potential $V(\varphi)$; $L_m$ is the Lagrangian of the perfect
isotropic fluid i.e. a fluid with only one radial pressure; $f(\varphi)$ is some function describing the
non-minimal interaction between the fluid and the scalar field.
The case $f=1$ corresponds to the absence of the non-minimal coupling,
but even in this case the two sources are still coupled via gravity.

We choose the Lagrangian for the isentropic  perfect fluid to have the form $L_m=p$ \cite{Stanuk1964,Stanuk}.
Using this Lagrangian, the corresponding energy-momentum tensor is (details are given in Appendix \ref{appen_emt})
 \begin{equation}
\label{emt_cham_star}
T_i^k=f\left[(\rho+p)u_i u^k-\delta_i^k p\right]+\partial_{i}\varphi\partial^{k}\varphi
-\delta_i^k\left[\frac{1}{2}\partial_{\mu}\varphi\partial^{\mu}\varphi-V(\varphi)\right] ~,
\end{equation}
where $\rho$ and $p$ are the density and the pressure of the fluid, $u^i$ is the four-velocity (here and throughout the paper
we set $c=1$). We take the static metric of the form
 \begin{equation}
\label{metric_sphera}
ds^2=e^{\nu(r)}dt^2-e^{\lambda(r)}dr^2-r^2d\Omega^2,
\end{equation}
where $d\Omega^2$ is the metric on the unit 2-sphere.
The $(_0^0)$ and $(_1^1)$ components of the Einstein equations for the metric
\eqref{metric_sphera} and the energy-momentum tensor \eqref{emt_cham_star} are
\begin{eqnarray}
\label{Einstein-00_cham_star}
&&G_0^0=-e^{-\lambda}\left(\frac{1}{r^2}-\frac{\lambda^\prime}{r}\right)+\frac{1}{r^2}
=8\pi G T_0^0=
8\pi G\left[f \rho+\frac{1}{2}e^{-\lambda}\varphi^{\prime 2}+V(\varphi)\right],
 \\
\label{Einstein-11_cham_star}
&&G_1^1=-e^{-\lambda}\left(\frac{1}{r^2}+\frac{\nu^\prime}{r}\right)+\frac{1}{r^2}
=8\pi G T_1^1=
8\pi G\left[-f p-\frac{1}{2}e^{-\lambda}\varphi^{\prime 2}+V(\varphi)\right].
\end{eqnarray}
The equation for the scalar field $\varphi$ coming from the Lagrangian \eqref{lagran_cham_star} is
$$
\frac{1}{\sqrt{-g}}\frac{\partial}{\partial x^i}\left[\sqrt{-g}g^{ik}\frac{\partial \varphi}{\partial x^k}\right]=
-\frac{d V}{d \varphi}+L_m \frac{d f}{d \varphi}.
$$
Using this field equation above with the perfect fluid $L_m=p$, and the metric
\eqref{metric_sphera}, gives the following scalar field equation
\begin{equation}
\label{sf_cham_star}
\varphi^{\prime\prime}+\left[\frac{2}{r}+\frac{1}{2}\left(\nu^\prime-\lambda^\prime\right)\right]\varphi^\prime=
e^{\lambda}\left(\frac{d V}{d \varphi}-p\frac{d f}{d\varphi}\right) ~,
\end{equation}
where the prime denotes  differentiation with respect to $r$. The Einstein field equations
are not all independent because of the relation $T^k_{i;k}=0$. The $i=1$ component of this equation has the form
\begin{equation}
\label{conserv_1_cham_star}
\frac{\partial T^1_1}{\partial r}+
\frac{1}{2}\left(T_1^1-T_0^0\right)\nu^\prime+\frac{2}{r}\left[T_1^1-\frac{1}{2}\left(T^2_2+T^3_3\right)\right]=0.
\end{equation}
Taking into account the expressions
$$
T_2^2=T_3^3=-f p+\frac{1}{2}e^{-\lambda}\varphi^{\prime 2}+V(\varphi) ~,
$$
and $T_0 ^0$, $T_1 ^1$ from \eqref{Einstein-00_cham_star}, \eqref{Einstein-11_cham_star} and using
\eqref{sf_cham_star}, allows us to write \eqref{conserv_1_cham_star} as
\begin{equation}
\label{conserv_2_cham_star}
\frac{d p}{d r}=-\frac{1}{2}(\rho+p)\frac{d\nu}{d r}.
\end{equation}
The matter Lagrangian used here, $L_m=p$, is not the only possibility. Other variants can be found
in \cite{Bertolami:2008ab}. However, the choice $L_m=p$ has the simplifying feature
\eqref{conserv_2_cham_star} does not contain additional terms involving the coupling
function $f (\varphi )$.

For a polytropic equation of state one has
\begin{equation}
\label{eqs_cham_star}
p=k \rho^\gamma,
\end{equation}
where $k, \gamma$ are constants. Now one can introduce the new variable
$\theta$ defined as \cite{Zeld}
\begin{equation}
\label{theta_def}
\rho=\rho_c \theta^n ~.
\end{equation}
Here $\rho_c$ is the central density, and the constant $n$, the polytropic index, is related to $\gamma$ via
$n=1/(\gamma-1)$. Putting these definitions together gives equation \eqref{eqs_cham_star} in the form
\begin{equation}
\label{eqs_fluid_theta}
p=k\,\rho^\gamma=k \rho^{1+1/n}=k\rho_c^{1+1/n} \theta^{n+1}.
\end{equation}
Using \eqref{eqs_fluid_theta} in equation \eqref{conserv_2_cham_star} leads to
\begin{equation}
\label{conserv_3_cham_star}
2\sigma(n+1)\frac{d\theta}{d r}=-(1+\sigma \theta)\frac{d\nu}{dr},
\end{equation}
with $\sigma=k \rho_c^{1/n}=p_c/\rho_c$ and $p_c$ is the pressure of the fluid at the center of the configuration.
This equation may be integrated to give $e^{\nu}$ in terms of $\theta$:
\begin{equation}
\label{nu_app}
	e^{\nu}=e^{\nu_c}\left(\frac{1+\sigma}{1+\sigma \theta}\right)^{2(n+1)},
\end{equation}
where $e^{\nu_c}$ is the value of $e^{\nu}$ at the center of the configuration where $\theta=1$.
The integration constant $\nu_c$, corresponds to the value of $\nu$ at the center of the
configuration. It is determined by requiring $e^{\nu}=1$ at infinity that i.e. that the
space-time is asymptotically flat.

The gravitating system of a real scalar field interacting with a perfect fluid
is characterized by three unknown functions -- $\lambda, \theta$ and $\varphi$. These three
functions are determined by the three equations \eqref{Einstein-00_cham_star}, \eqref{Einstein-11_cham_star}
and \eqref{sf_cham_star}, and also by the relation \eqref{nu_app}. We now rewrite these equations
by the introduction of a new function $u(r)$ \cite{Tooper:1964}
\begin{equation}
\label{u_app}
u(r)=\frac{r}{2 G M}\left(1-e^{-\lambda}\right) \rightarrow e^{-\lambda}=1-\frac{2 G M u}{r}.
\end{equation}
Here $M$ is the mass of the configuration within the range $0 \leq r \leq r_b$,
where $r_b$ is the boundary of the fluid where $\theta=0$.
Using this function, equation \eqref{Einstein-00_cham_star} becomes
\begin{equation}
\label{00_via_u}
M\frac{d u}{d r}=4\pi r^2\left[f\rho+\frac{1}{2}\left(1-\frac{2 G M u}{r}\right)\varphi^{\prime 2}+V\right].
\end{equation}
From \eqref{u_app} one can define $M(r) \equiv M u(r)$ which can be interpreted as the total mass of the configuration
in the range $[0, r]$. This mass has contributions from both the fluid and the scalar field, within a sphere of
coordinate radius $r$. To avoid a singularity in $M(r)$ at the origin, one has to put $u(0)=0$  \cite{Tooper:1964}.
This corresponds to the fact that the mass at the origin is equal to zero i.e. $M(0) =0$.

In anticipation of analyzing the system of equations -- \eqref{Einstein-00_cham_star}, \eqref{Einstein-11_cham_star}
and \eqref{sf_cham_star}) -- numerically, we introduce the following dimensionless variables
\begin{equation}
\label{dimless_xi_v}
\xi=A r, \quad v(\xi)=\frac{A^3 M}{4\pi \rho_c} u(r),
\quad \phi(\xi)=\left[\frac{4\pi G}{\sigma(n+1)}\right]^{1/2}\varphi(r),
\quad \text{where} \quad A=\left[\frac{4\pi G \rho_c}{(n+1)k \rho_c^{1/n}}\right]^{1/2},
\end{equation}
$A$ has the dimensions of an inverse length. With this one can rewrite equations
\eqref{Einstein-00_cham_star} and \eqref{Einstein-11_cham_star} in the form
\begin{eqnarray}
\label{eq_v_app}
\frac{d v}{d\xi} &=& \xi^2\left\{f\theta^n+\frac{1}{2}\left[
1-2\sigma(n+1)\frac{v}{\xi}
\right]
\left(\frac{d\phi}{d\xi}\right)^2+\tilde{V}\right\}
,\\
\label{eq_theta_app}
\xi^2\frac{1-\frac{2\sigma(n+1)v}{\xi}}{1+\sigma\theta}\frac{d\theta}{d\xi} &=&
\xi^3\left[f\theta^n\left(1-\sigma\theta\right)+
2 \tilde{V}-\frac{1}{\xi^2}\frac{d v}{d\xi}\right]-v ~,
\end{eqnarray}
where $\tilde{V}=V/\rho_c$ is the dimensionless potential energy of the field.

Next, using  \eqref{conserv_3_cham_star}, one can rewrite equation \eqref{sf_cham_star} as follows:
\begin{eqnarray}
\label{eq_phi_dim_cham_star}
\frac{d^2 \phi}{d\xi^2}&+&\left\{\frac{2}{\xi}-\frac{\sigma(n+1)}{1+\sigma \theta}
\left[\frac{d\theta}{d\xi}+\frac{1+\sigma \theta}{1-\frac{2\sigma(n+1)v}{\xi}}\frac{1}{\xi}
\left(\frac{d v}{d\xi}-\frac{v}{\xi}\right)\right]\right\}\frac{d\phi}{d\xi}= \nonumber\\
&&\left[1-2\sigma(n+1)\frac{v}{\xi}\right]^{-1}\left(\frac{d \tilde{V}}{d\phi}-\sigma \theta^{n+1}\frac{d f}{d\phi}\right).
\end{eqnarray}
Thus the static configuration under consideration is described by the three equations \eqref{eq_v_app}-\eqref{eq_phi_dim_cham_star}.

\section{Configuration with a mass and a quartic self-interaction term}
\label{static_sol_cham_star}

In this section we will show that there are non-singular, finite-mass solutions of equations
\eqref{eq_v_app}-\eqref{eq_phi_dim_cham_star}. First we specify the boundary conditions.
Using the above dimensionless variables we ``normalize" $\theta$ to unity at the center of the configuration $\xi=0$
\begin{equation}
\label{bound_theta}
\theta_0\equiv\theta(0)=1.
\end{equation}
From equation \eqref{eq_v_app} one can show that $v\to 0$ like $\xi^3$ as $\xi\to 0$. Combining this with
equation \eqref{eq_theta_app} one in turn finds $d\theta/d\xi \to 0$ as $\xi \to 0$.
Bearing in mind that we are looking for regular solutions, we define the boundary conditions in the vicinity of
$\xi \approx 0$ as
\begin{equation}
\label{bound_all}
\theta \approx \theta_0+\frac{\theta_2}{2}\xi^2, \quad v \approx v_3 \xi^3, \quad
\phi \approx \phi_0+\frac{\phi_2}{2}\xi^2,
\end{equation}
where  $\phi_0$ corresponds to the initial value of the scalar field
$\phi$, the parameters $\theta_2, v_3$  are arbitrary, and the value of the coefficient
$\phi_2$ is defined from equation \eqref{eq_phi_dim_cham_star} as
$$
\phi_2=\frac{1}{3}\left[
	\left(\frac{d \tilde{V}}{d\phi}
	\right)_0-\sigma \theta_0^{n+1}\left(\frac{d f}{d\phi}\right)_0
\right].
$$
The index $0$ denotes that the values of the functions are taken at $\xi=0$.

Using the boundary conditions \eqref{bound_all}, we proceed to solve the system
\eqref{eq_v_app}-\eqref{eq_phi_dim_cham_star} numerically. The behavior of the solution will depend both on
the parameters of the polytropic fluid $n, \sigma$ and the form of the potential energy of the scalar field
$\tilde{V}$, and the coupling function  $f$. One of the simplest and most commonly used choices for the potential energy
is that of a scalar field with mass $m$ and a quartic self-interaction term ($\kappa>0$)
$$
V=\frac{1}{2} m^2 \varphi^2+\frac{1}{4} \kappa \varphi^4.
$$
It was pointed out in \cite{jetzer} that the system with such potential has only singular
static solutions. Below we show that an inclusion of the non-minimal interaction between the
scalar field and the polytropic fluid allows one to obtain regular static solutions.
As an example, let us choose the coupling function  $f$ in dimensionless form as follows
\begin{equation}
\label{f_def}
f=\frac{\beta}{2}\phi^2, \quad \beta>0.
\end{equation}
Then, using the dimensionless variables introduced in the previous section, the potential
can be rewritten as
\begin{equation}
\label{poten_dim}
\tilde{V}=\frac{1}{2} \mu^2 \phi^2+\frac{1}{4} \Lambda \phi^4,
\end{equation}
where the new dimensionless constants are given as
$$
\mu=\frac{m^2}{\rho_c}\frac{\sigma (n+1)}{4\pi G}, \quad \Lambda=\frac{\kappa}{\rho_c}\left[\frac{\sigma (n+1)}{4\pi G}\right]^2 .
$$
Using the above potential $\tilde{V}$ and the function $f$, equations
 \eqref{eq_v_app}-\eqref{eq_phi_dim_cham_star} take the form
\begin{eqnarray}
\label{eq_theta_app_n}
&&\xi^2\frac{1-2\sigma(n+1)v/\xi}{1+\sigma\theta}\frac{d\theta}{d\xi}=
\xi^3\left[\frac{\beta}{2}\phi^2\theta^n\left(1-\sigma\theta\right)+
 \mu^2 \phi^2+\frac{1}{2} \Lambda \phi^4-\frac{1}{\xi^2}\frac{d v}{d\xi}\right]-v
,\\
\label{eq_v_app_n}
&&\frac{d v}{d\xi}=\frac{\xi^2}{2}\left\{\beta\phi^2\theta^n+
\left[1-2\sigma(n+1)\frac{v}{\xi}\right]\left(\frac{d\phi}{d\xi}\right)^2+ \mu^2 \phi^2+\frac{1}{2} \Lambda \phi^4\right\},\\
\label{eq_phi_dim_cham_star_n}
\frac{d^2 \phi}{d\xi^2}&+&\left\{\frac{2}{\xi}-\frac{\sigma(n+1)}{1+\sigma \theta}
\left[\frac{d\theta}{d\xi}+\frac{1+\sigma \theta}{1-\frac{2\sigma(n+1)v}{\xi}}\frac{1}{\xi}
\left(\frac{d v}{d\xi}-\frac{v}{\xi}\right)\right]\right\}\frac{d\phi}{d\xi}= \nonumber\\
&&\left[1-2\sigma(n+1)\frac{v}{\xi}\right]^{-1}
\left[\left(\mu^2-\beta \sigma\theta^{n+1}\right)\phi+\Lambda \phi^3\right].
\end{eqnarray}

The parameter $\mu$ can be absorbed by introducing the rescalings $x=\mu\xi$, $\bar{v}=\mu v$,
$\bar{\beta}=\beta/\mu^2$, $\bar{\Lambda}=\Lambda/\mu^2$. Bearing this in mind, we assume $\mu=1$ in further calculations.
One can see from these equations that the presence of the interaction between the scalar field and the fluid is defined by the
parameter $\beta$. In the absence of the fluid, this system only has singular solutions \cite{jetzer}.
The inclusion of the fluid changes the situation since the presence of the term  $\beta \sigma\theta^{n+1}$
on the right hand side of  \eqref{eq_phi_dim_cham_star_n} for the scalar field corresponds to an effective mass term
$\left(\mu^2-\beta \sigma\theta^{n+1}\right)$ whose sign depends both on  the behavior of the fluid density
$\theta$ and the values of the parameters $\beta$ and $\sigma$. We note that the present system of gravitating real scalar field
plus perfect fluid turns out to be similar to the gravitating complex scalar scalar field system considered in \cite{Colpi}.
The gravitating complex scalar field studied in \cite{Colpi} had a potential of the type \eqref{poten_dim}
and it gave regular, {\it stationary} solutions. Thus this is already a hint that we can expect regular, stationary
solutions for the present system.

\section{Numerical results}
\label{num_calc}

\subsection{Relativistic case}

The results of the numerical calculations for the system
\eqref{eq_theta_app_n}-\eqref{eq_phi_dim_cham_star_n} with the boundary conditions \eqref{bound_all} are presented in tables
\ref{tab1} and \ref{tab2}. The solutions were started near the origin (i.e. near $\xi \approx 0$) and solved out to a point
$\xi = \xi_1$  where the function $\theta$ becomes zero. From equations \eqref{eqs_cham_star} and \eqref{theta_def}
this is where the fluid vanishes and it is this point, $\xi = \xi_1$, that we define to be the surface of the star.
Beyond the point $\xi=\xi_1$ we continued the numerical solutions with only the gravitational and scalar fields
while the fluid was set to zero. The interior and exterior solutions were then connected to one another. Details of this are
given below.  Previous numerical studies with a fluid having a polytropic equation of state \cite{Tooper:1964}
found regular, relativistic star-like configurations. These star-like solutions of \cite{Tooper:1964}
where found for the values of the parameters $0 \leq \sigma \leq 0.75$  and $1 \leq n \leq 3$.
Since our system has two additional parameters ($\Lambda, \beta$) we restrict ourselves to examine just two sets of the parameters
(i.e. $\sigma=0.2, n=1.0$ and $\sigma=0.2, n=1.5$) at different values of $\Lambda$ and $\beta$ when looking for regular solutions.
Below we show that these two sets of solutions differ considerably in the behavior of their physical characteristics.

The parameters of the system under consideration for the chosen values of $\sigma$ and $n$ are given
in tables \ref{tab1} and \ref{tab2}. The procedure for finding these parameters is the following: given the values of
$\sigma$ and $n$, we seek {\it eigenvalues} (i.e. values of the parameters $\Lambda$ and $\beta$) for which
the function $\theta$ goes to zero at some finite value of  $\xi = \xi_1$ which as mentioned above we take
to correspond to the surface of the star. Then since $M(r) = M u(r)$ (where $M$ is the total mass
of the star between $r=0$ and its surface at $r =r_b$) it is required that $u(r_b) =1$.
Using this we can evaluate the function $v(\xi)$ from \eqref{dimless_xi_v} at $\xi = \xi_1$
\begin{equation}
\label{v_bound}
v(\xi_1)=\frac{A^3 M}{4\pi \rho_c}.
\end{equation}
This quantity defines the total mass $M$ of the configuration through the parameters of the fluid
$\sigma$, $n$ and $\rho_c$ which determine the parameter $A$.

The value $\xi_1$ of the coordinate $\xi$, corresponding to the boundary of the configuration, does not represent the
radius of the star as measured by a distant observer. To define this radius it is necessary to make a coordinate
transformation to a new dimensionless variable $\bar{\xi}$ which is defined as follows
$$
\bar{\xi}=\int_0^{\xi} e^{\lambda/2}d\xi,
$$
or taking into account
 \eqref{u_app} and \eqref{dimless_xi_v}
\begin{equation}
\label{xi_observ}
\bar{\xi}=\int_0^{\xi} \left[1-2\sigma(n+1)v(\xi)/\xi\right]^{-1/2}d\xi.
\end{equation}
Then the observable radius of the configuration $\bar{R}$ in dimensional variables is defined as $\bar{R}=\bar{\xi_1}/A$
in accordance with the data presented in tables \ref{tab1} and \ref{tab2}.

\begin{table}[htbp]
 \caption{The parameters of the relativistic configurations with the constant central value of the scalar field
 $\phi_0=0.45$ }
\begin{center}
\begin{tabular}{|p{2cm}|p{2cm}|p{2cm}|p{2cm}|p{2cm}|p{2cm}|p{2cm}|}
\hline
$\Lambda$ & $\beta$ & $\xi_1$ & $\bar{\xi_1}$& $v(\xi_1)$&$\rho_c/\bar{\rho}$&$-\Omega/M$\\
\hline
\multicolumn{7}{|c|}{$\sigma=0.2, n=1.5$}\\
\hline
1500&	10200& 	0.37&	0.4017&	  0.0269&	0.6273&		0.1344\\
1250&	8500&	0.41&	0.4450&	  0.0295&	0.7786&		0.1344\\
1000&	6900&	0.47&	0.5092&	  0.0326&	1.0609&		0.1341\\
750	&   5200&	0.54&	0.5851&	  0.0376&	1.3972&		0.1341\\
500& 	3500&	0.66&	0.7150&	  0.0458&	2.0939&		0.1340\\
250& 	1800&	0.95&	1.0277&	  0.0637&	4.4834&		0.1337\\
150& 	1130&	1.25&	1.3494&	  0.0801&	8.1296&		0.1331\\
100& 	790&   	1.58&	1.7012&	  0.0955&	13.7681&		0.1326\\
75&  	620&	1.83&	1.9678&	  0.1075&	19.0029&		0.1322\\
60&  	525&    2.13&	2.2830&	  0.1160&	27.7771&		0.1316\\
41&  	395&    2.70&   2.8820&	  0.1330&	49.3308&		0.1308\\
30&  	325&    3.35&	3.5573&	  0.1449&	86.4977&		0.1299\\
20&  	260.5&	4.70&  	4.9472&	  0.1594&	217.0848&		0.1287\\
10&  	195.24&	9.00&    9.3259&	  0.1798&	1351.2011&		0.1269\\
5&   	162.1683&	21.00&	21.4257&	0.1983&	15568.1073&		0.1229\\
3&   	148.8345216&	30.00&	30.4723&	0.2102&	42819.2173&		0.1192\\
\hline
\multicolumn{7}{|c|}{$\sigma=0.2, n=1.0$}\\
\hline
1500.0&	8700&   	0.210&	0.2315&	0.0313&	0.0985&		0.1229\\
1250.0&	7250&	    0.230&	0.2535&	0.0343&	0.1181&		0.1229\\
1000.0&	5800&	    0.257&	0.2833&	0.0384&	0.1473&		0.1230\\
500.0&    2910&	    0.362&	0.3992&	0.0543&	0.2909&		0.1231\\
350.0& 	2050&	    0.430&	0.4742&	0.0648&	0.4090&		0.1231\\
200.0& 	1200&	    0.567&	0.6250&	0.0846&	0.7181&		0.1229\\
150.0& 	920&    	0.655&	0.7216&	0.0965&	0.9712&		0.1227\\
100.0& 	640&    	0.800&	0.8805&	0.1152&	1.4810&		0.1222\\
50.0&  	350&    	1.110&	1.2204&	0.1559&	2.9250&		0.1217\\
30.0&  	238&    	1.400&	1.5362&	0.1880&	4.8644&		0.1209\\
20.0&  	180&    	1.650&	1.8084&	0.2160&	6.9318&		0.1203\\
10.0&  	126&    	2.160&	2.3567&	0.2551&	13.1668&		0.1185\\
6.8& 	109&    	2.442&	2.6577&	0.2724&	17.8189&		0.1175\\
5.0& 	100&    	2.670&	2.8992&	0.2825&	22.4626&		0.1167\\
3.0& 	90&     	3.000&	3.2474&	0.2950&	30.5069&		0.1157\\
1.0& 	80&     	3.500&	3.7719&	0.3094&	46.1887&		0.1145\\
0.3& 	76&     	3.670&	3.9511&	0.3169&	52.0002&		0.1141\\
0.1& 	75&     	3.750&	4.0345&	0.3186&	55.1796&		0.1140\\
\hline
\end{tabular}
\end{center}
\label{tab1}
\end{table}

\begin{table}[htbp]
 \caption{The parameters of the relativistic configurations as a function of the central value of the scalar field
 $\phi_0$ for fixed observable radius of the star $\bar{\xi}_1$ and for given values of $\sigma$ and $n$.}
\begin{center}
\begin{tabular}{|p{2cm}|p{2cm}|p{2cm}|p{2cm}|p{2cm}|p{2cm}|p{2cm}|}
\hline
$\phi_0$&$\Lambda$ & $\beta$ & $\xi_1$ & $v(\xi_1)$&$\rho_c/\bar{\rho}$&$-\Omega/M$\\
\hline
\multicolumn{7}{|c|}{$\sigma=0.2, n=1.5$,  $\bar{\xi}_1=3.219$}\\
\hline
\multicolumn{7}{|c|}{\bf Relativistic star without a scalar field }\\
\hline
&   &    &    2.699&	0.9604&	6.8270&		0.2207\\
\hline
\multicolumn{7}{|c|}{\bf Relativistic star with the scalar field}\\
\hline
0.30&	2101.0&	1642.5&	3.055&	0.1022&	92.9763&		0.1325\\
0.35&	617.0&	868.0&	3.040&	0.1186&	78.9467&		0.1323\\
0.40&	184.0&	531.0&	3.030&	0.1306&	70.9796&		0.1316\\
0.44&	54.5&	385.0&	3.025&	0.1376&	67.0397&		0.1308\\
0.45&	34.9&	355.0&	3.025&	0.1396&	66.0795&		0.1304\\
0.47&	6.6&    317.0&	3.025&	0.1402&	65.8043&		0.1297\\
\hline
\multicolumn{7}{|c|}{$\sigma=0.2, n=1.0$,  $\bar{\xi}_1=2.657$}\\
\hline
\multicolumn{7}{|c|}{\bf Relativistic star without a scalar field }\\
\hline
&   &    &    2.277&	1.1430&	3.4430&		0.2108\\
\hline
\multicolumn{7}{|c|}{\bf Relativistic star with the scalar field}\\
\hline
0.27&	510.15&	275.10&	2.424&	0.2986&	15.8983&		0.1262\\
0.30&	300.40&	225.90&	2.425&	0.2947&	16.1309&		0.1250\\
0.33&	178.90&	190.30&	2.427&	0.2903&	16.4159&		0.1237\\
0.35&	124.50&	169.95&	2.430&	0.2889&	16.5532&		0.1228\\
0.39&	57.70&	142.00&	2.436&	0.2794&	17.2429&		0.1207\\
0.41&	36.00&	130.00&	2.442&	0.2769&	17.5278&		0.1197\\
0.43&	19.30&	119.00&	2.442&	0.2739&	17.7255&		0.1185\\
0.45&	6.80&	109.00&	2.442&	0.2724&	17.8189&		0.1175\\
0.46&	1.90&	105.00&	2.442&	0.2714&	17.8864&		0.1171\\
\hline
\end{tabular}
\end{center}
\label{tab2}
\end{table}

\begin{figure}[t]
\begin{minipage}[t]{.49\linewidth}
  \begin{center}
  \includegraphics[width=9.5cm]{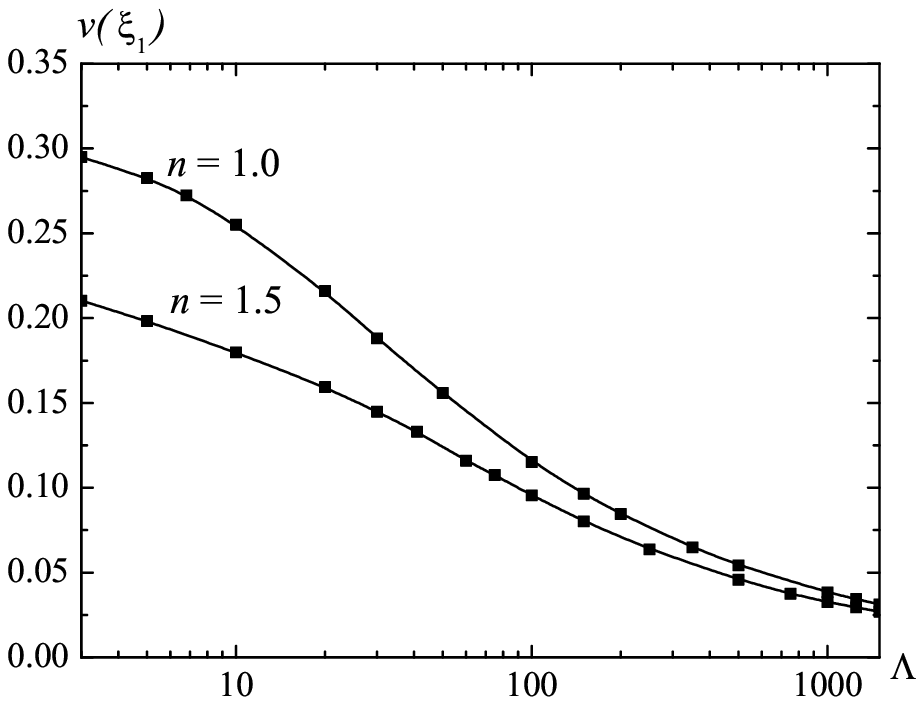}
\vspace{-1.cm}
  \caption{\small The dependence of the function $v(\xi_1)$
  (which is proportional to the total mass $M$)
  on the value of the coupling parameter $\Lambda$ from \eqref{poten_dim}
  at the fixed central value $\phi_0=0.45$ and  $\sigma=0.2$. The data are taken from table \ref{tab1}.}
    \label{mass_lambda_fig}
  \end{center}
\end{minipage}\hfill
\begin{minipage}[t]{.49\linewidth}
  \begin{center}
  \includegraphics[width=9.5cm]{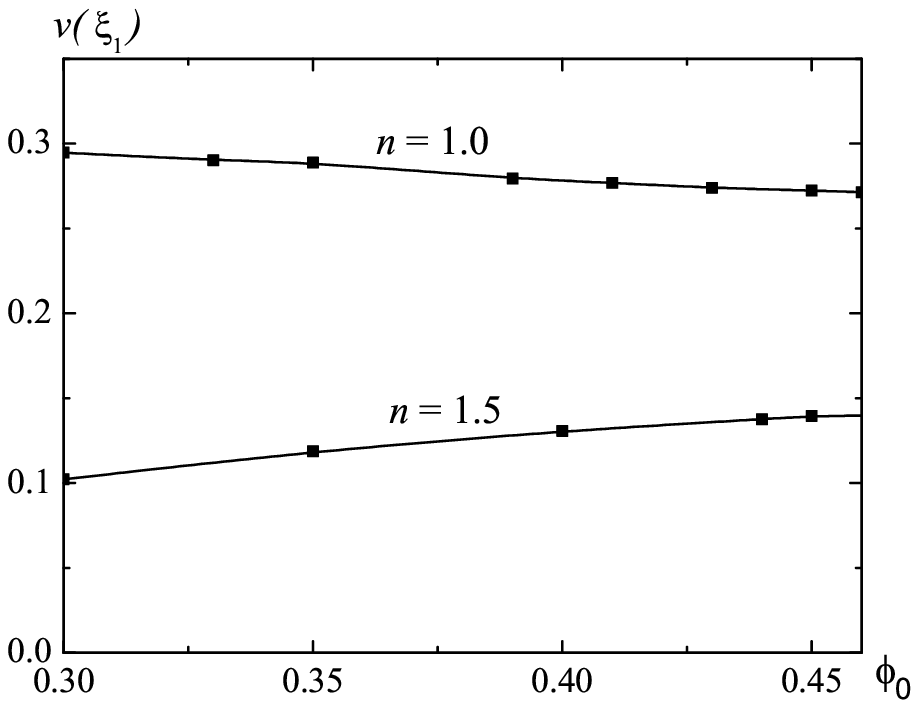}
\vspace{-1.1cm}
  \caption{\small
  The dependence of the function $v(\xi_1)$
  on the central value of  $\phi_0$ at the fixed  $\bar{\xi}_1=3.219$ for $n=1.5$
  and $\bar{\xi}_1=2.657$ for $n=1.0$, and $\sigma=0.2$ for both graphs.
The data are taken from table \ref{tab2}.}
    \label{mass_phi_fig}
  \end{center}
\end{minipage}\hfill
\end{figure}

\begin{figure}[t]
\begin{minipage}[t]{.49\linewidth}
  \begin{center}
  \includegraphics[width=7cm]{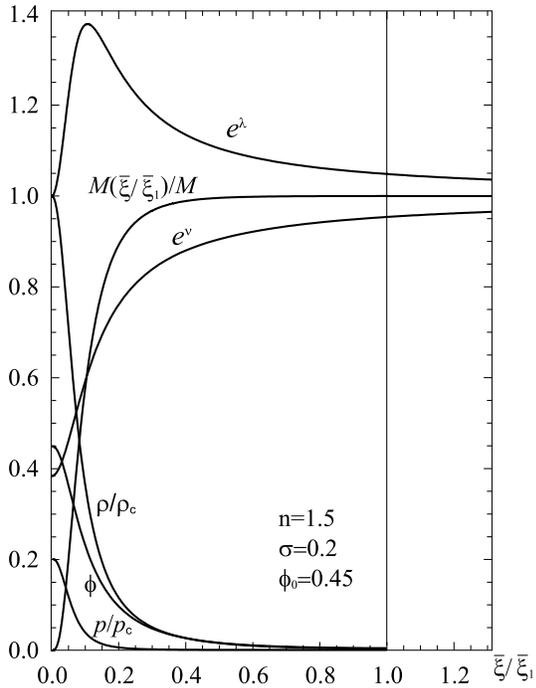}
  \end{center}
\end{minipage}\hfill
\begin{minipage}[t]{.49\linewidth}
  \begin{center}
  \includegraphics[width=7cm]{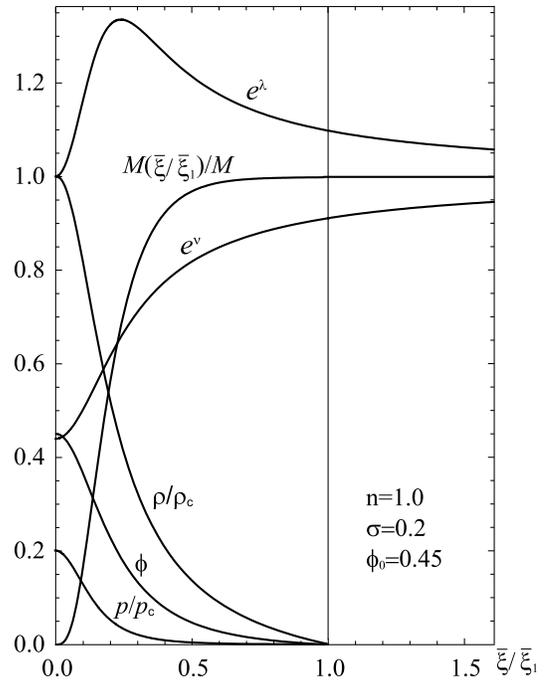}
  \end{center}
\end{minipage}\hfill
\caption{\small The density $\rho$ and the pressure $p$ of the fluid, the scalar field $\phi$,
the metric functions $e^\nu$, $e^\lambda$ and the current mass $M$
  as functions of the relative invariant radius $\bar{\xi}/\bar{\xi_1}$ for $n=1.5$ and
$n=1.0$ respectively. To provide the asymptotic flatness of the solutions,
i.e.~$e^\nu, e^\lambda \to 1$ as $\bar{\xi} \to \infty$,
the value of the constant $\nu_c$ needed to be chosen as
 $\nu_c \approx -0.958$ for $n=1.5$ and $\nu_c \approx -0.822$ for $n=1.0$.}
    \label{metr_figs}
\end{figure}

\begin{figure}[t]
\begin{minipage}[t]{.49\linewidth}
  \begin{center}
  \includegraphics[width=7cm]{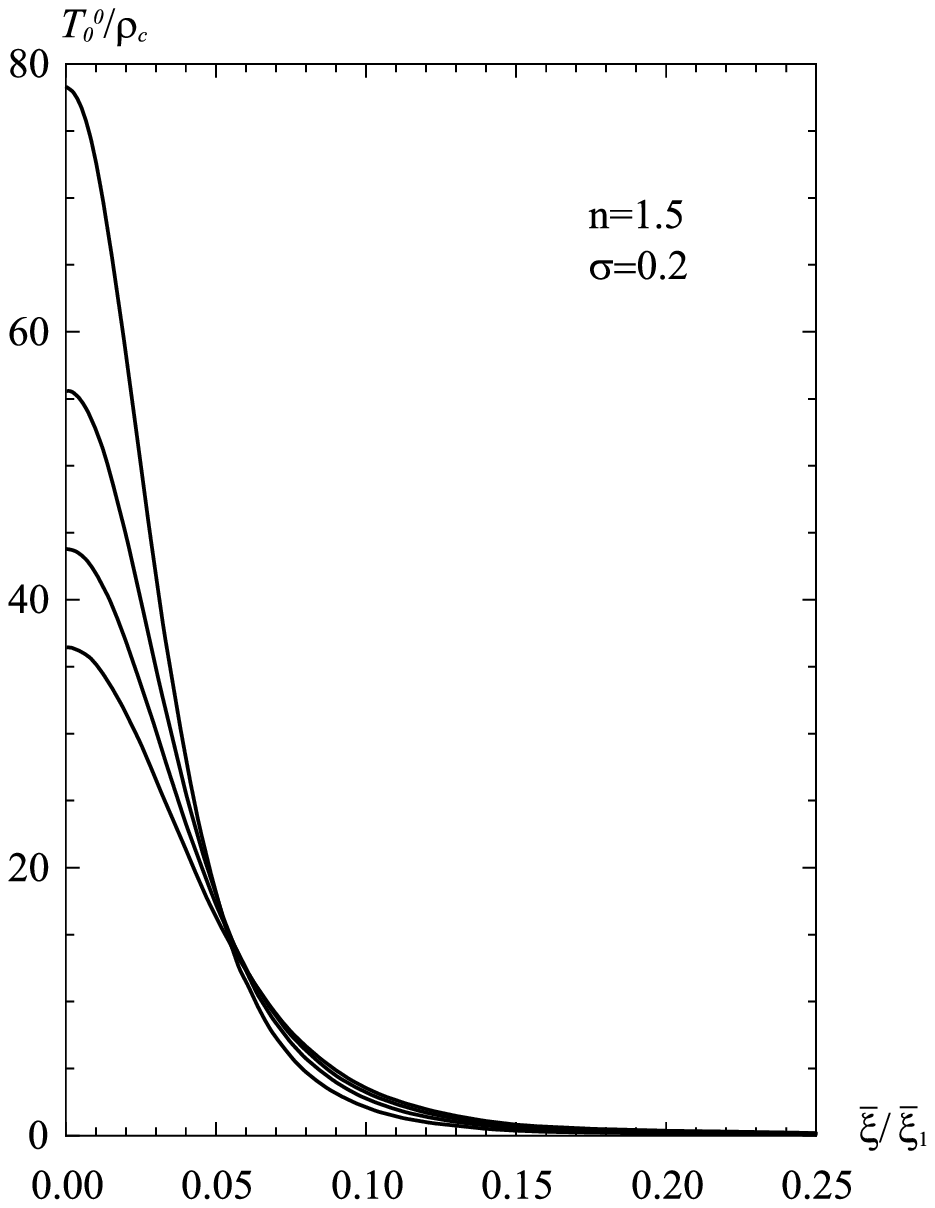}
  \caption{\small The distributions of the total energy density in units of $\rho_c$ from \eqref{tot_dens_poly_scalar}
  for different the initial values of $\phi_0=0.30, 0.35, 0.40, 0.45$, from top to bottom.}
    \label{energ_n_1_5_fig}
  \end{center}
\end{minipage}\hfill
\begin{minipage}[t]{.49\linewidth}
  \begin{center}
  \includegraphics[width=7cm]{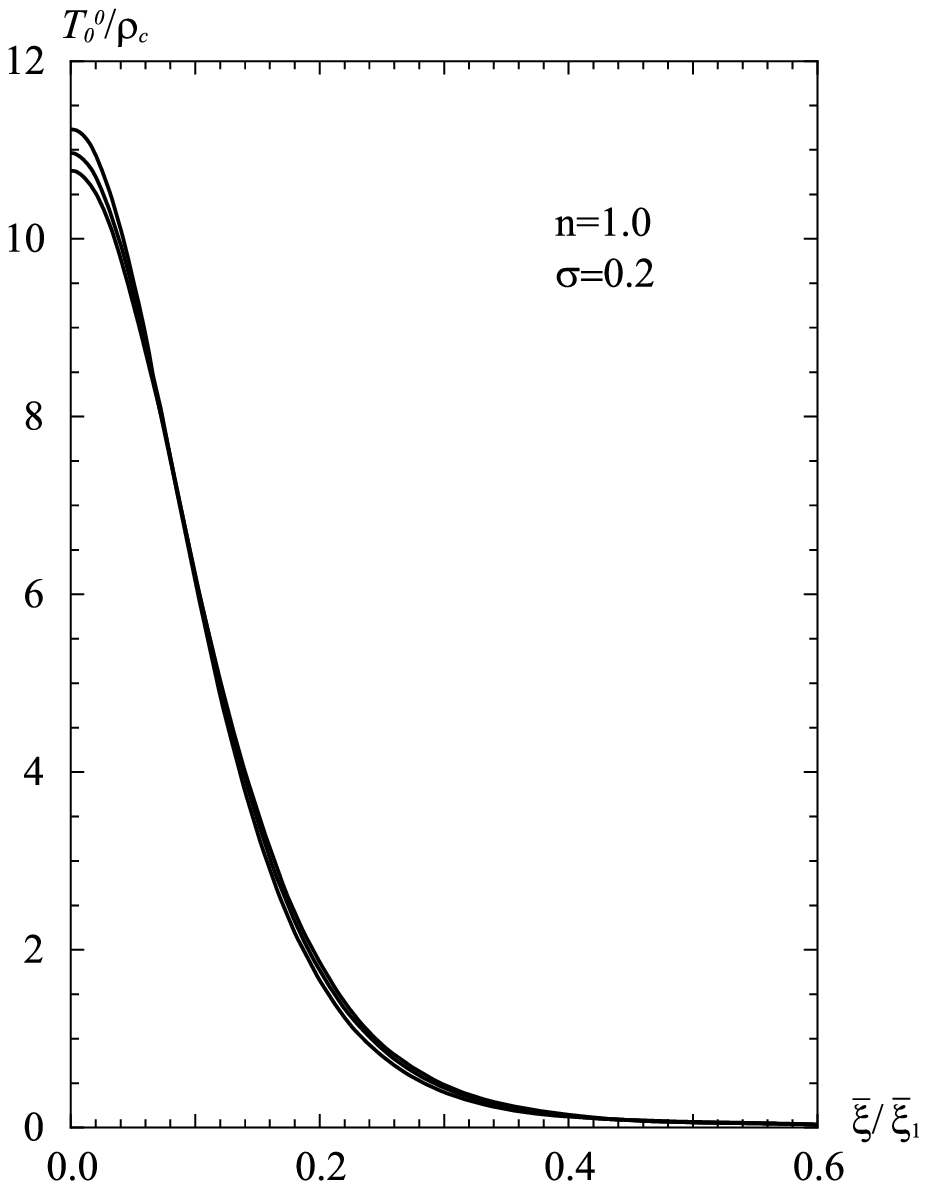}
  \caption{\small The distributions of the total energy density in units of $\rho_c$ from \eqref{tot_dens_poly_scalar}
  for the different initial values of $\phi_0=0.45, 0.35, 0.27$, from top to bottom.}
 \label{energ_n_1_0_fig}
  \end{center}
\end{minipage}\hfill
\end{figure}

Also the ratio of the central density $\rho_c$ (of the fluid only)
to the average density  $\bar{\rho}$ (of both the fluid and
the scalar field) is presented in the tables. We define the average density as \cite{Tooper:1964}
$$
\bar{\rho}=\frac{M}{(4/3)\pi R^3}=\frac{3 M A^3}{4\pi \xi_1^3}.
$$
Using the expression for $v(\xi)$ from  \eqref{v_bound}, one can express the quantity
$M A^3/4\pi$ in terms of the central density $\rho_c$ and the boundary value $v(\xi_1)$ of the mass function. Then we obtain the
relation
$$
\frac{\rho_c}{\bar{\rho}}=\frac{\xi_1^3}{3 v(\xi_1)}.
$$

Finally, in the last column of both tables the gravitational potential energy expressed in
units of $M$  is shown (see \cite{Tooper:1964} for more details):
$$
\frac{\Omega}{M}=1-\frac{1}{v(\xi_1)}\int_0^{\xi_1} \frac{T_0^0 \xi^2 d\xi}{\left[1-2\sigma(n+1)v(\xi)/\xi\right]^{1/2}},
$$
where the expression for the energy density is defined from \eqref{emt_cham_star}. Using
equations \eqref{u_app},  \eqref{dimless_xi_v}, \eqref{f_def} and \eqref{poten_dim} one can write
the energy density in the following form (in units of $\rho_c$)
\begin{equation}
\label{tot_dens_poly_scalar}
T_0^0=\frac{1}{2}\left\{\beta\phi^2\theta^n+
\left[1-2\sigma(n+1)\frac{v}{\xi}\right]\left(\frac{d\phi}{d\xi}\right)^2+ \mu^2 \phi^2+\frac{1}{2} \Lambda \phi^4\right\}.
\end{equation}
The absolute value of the potential energy $\Omega$ represents the work that would have to be done on the system to diffuse its mass
to infinity.

Next, using the data from the tables, in figure \ref{mass_lambda_fig} we plot
$v(\xi_1)$, which is proportional to the total mass of the configuration
$M$, on the value of the self-coupling parameter  $\Lambda$
at some constant central  $\phi_0=0.45$. In figure \ref{mass_phi_fig} we plot $v(\xi_1)$
as a function of the value of central $\phi_0$ for different $n$ values and $\sigma =0.2$.
From figure \ref{mass_lambda_fig} one can see that, while $\Lambda$ increases, the masses of the configurations decrease
both for $n=1.0$ and $n=1.5$. Initially, for small $\Lambda$, the masses are very different for $n=1.0$ and $n=1.5$; asymptotically,
for large $\Lambda$, they approach comparable values. This is because
for large  $\Lambda$ the total mass is defined by the scalar field but not by the contribution from the fluid.
The sizes of the configurations are different for the two cases $n=1.0$ (which has a size characterized by $\bar{\xi}=2.657$)
and $n=1.5$ (which has a size characterized by $\bar{\xi}=3.219$).

On the other hand if one fixes the size of the star to equal the size of a corresponding relativistic configuration
without a scalar  field, and changes the central value of $\phi_0$ so as
to get regular solutions, it is necessary to choose appropriate eigenvalues of
$\Lambda$ and $\beta$. In this case $v(\xi_1)$ as a function of $\phi_0$ is given in figure \ref{mass_phi_fig}.
From this figure one sees that as $\phi_0$ increases the mass of the configuration can either
increase (for the $n=1.5$ case) or decreases (for the $n=1.0$ case).

It follows from tables \ref{tab1}, \ref{tab2} and figures \ref{mass_lambda_fig}, \ref{mass_phi_fig} that the masses
of the configurations, computed for the values of the system parameters used here,
are considerably smaller than the masses of the stars  without a scalar field.
The maximal dimensionless masses of the configuration {\it with} the scalar field for $n=1.0$ and at $n=1.5$ are
$0.32$ and $0.21$, respectively; {\it without} a scalar field the masses of the same configurations are
$1.14$ and $0.96$, respectively. Our attempts to increase the masses of the system by varying $\phi_0$ were not successful.
From figure \ref{mass_phi_fig} the implication is that the masses change only slightly with $\phi_0$.

Next, in figure \ref{metr_figs} the metric functions $e^{\lambda}, e^{\nu}$, and the mass distribution
$M(\bar{\xi}/\bar{\xi_1})$ (as a function of the dimensionless radius $\bar{\xi}$) are
given. In plotting these functions we used the following procedure:

{\it Inside the Star}: The function $e^{\nu}$ was plotted using \eqref{nu_app};
the function $e^\lambda$ was plotted from equation \eqref{u_app}, in terms of the dimensionless
variables from \eqref{dimless_xi_v}. Explicitly
\begin{equation}
\label{lambda_dim}
e^\lambda=\left[1-2\sigma(n+1)\frac{v}{\xi}\right]^{-1}.
\end{equation}
From the interior solution parts of figure \ref{metr_figs} (i.e. the region $0<\xi < \xi_1$) we see the reason for the
terminology ``chameleon star" for the present solutions -- the scalar field mimics the behavior of the fluid in the interior
region. Even in the exterior region where the fluid goes to zero the scalar field is asymptotically going to zero.

{\it Outside the star}: The solution goes to the Schwarzschild solution
\begin{equation}
\label{lambda_dim_ext}
e^{\nu}=e^{-\lambda}=1-2\sigma(n+1)\frac{v(\xi_1)}{\xi}.
\end{equation}
The mass function is defined as
$$
M(\bar{\xi}/\bar{\xi_1})=M\frac{v(\xi)}{v(\xi_1)}.
$$

Finally, in figures  \ref{energ_n_1_5_fig} and \ref{energ_n_1_0_fig} we plot the energy density, $T^0_0$,
from the expression \eqref{tot_dens_poly_scalar}.  The definition of the mass function given just above
takes into account the energy of the fluid and the scalar field from $\xi =0$ to $\xi = \xi_1$. Although the
fluid vanishes at $\xi =\xi_1$ the scalar field does not vanish. Thus in principle one should include the
contribution to the mass function of the scalar field from $\xi = \xi _1$ to $\xi \to \infty$. However from the
figures \ref{metr_figs}-\ref{energ_n_1_0_fig} one can see that $\phi$ and $\phi {\prime}$ rapidly
go to zero as $\xi \rightarrow \infty$ and thus the mass function is given essentially just by the fluid and
scalar field energy density between $\xi =0$ and $\xi = \xi_1$.

We now turn the the question of connecting the interior region (i.e. $0< \xi < \xi_1$ where one has both
fluid plus scalar field as a source) with the exterior region (i.e. $\xi \ge \xi_1$ where one has only a
rapidly vanishing scalar field). For this purpose, we write down the Einstein equations of
\eqref{Einstein-00_cham_star} and \eqref{Einstein-11_cham_star},
and the scalar field equation of \eqref{sf_cham_star}, without the fluid source i.e.~$\theta=0$ so that from
\eqref{eqs_cham_star} and \eqref{theta_def} we have zero pressure and density.
This leads to the following system of equations:
\begin{eqnarray}
\label{Einstein-00_cham_star_ext}
&&-e^{-\lambda}\left(\frac{1}{r^2}-\frac{\lambda^\prime}{r}\right)+\frac{1}{r^2}
=
8\pi G\left[\frac{1}{2}e^{-\lambda}\varphi^{\prime 2}+V(\varphi)\right],
 \\
\label{Einstein-11_cham_star_ext}
&&-e^{-\lambda}\left(\frac{1}{r^2}+\frac{\nu^\prime}{r}\right)+\frac{1}{r^2}
=
8\pi G\left[-\frac{1}{2}e^{-\lambda}\varphi^{\prime 2}+V(\varphi)\right],\\
\label{sf_eq_ext}
&&\varphi^{\prime \prime}+\left[\frac{2}{r}+\frac{1}{2}\left(\nu^\prime-\lambda^\prime\right)\right]\varphi^\prime=
e^\lambda \frac{d V}{d\varphi},
\end{eqnarray}
which, by analogy with the transformations made above,
can be rewritten in terms of the dimensionless variables
$v(\xi), \phi(\xi)$ and $\nu(\xi)$ as follows
\begin{eqnarray}
\label{eq_v_ext}
&&	\frac{d v}{d\xi} = \xi^2 \left[\frac{1}{2}e^{-\lambda}\left(\frac{d\phi}{d\xi}\right)^2+\tilde{V}\right],
 \\
\label{eq_nu_ext}
&&	\frac{d\nu}{d\xi} =
\frac{2\sigma(n+1)e^{\lambda}}{\xi}\left[\frac{v}{\xi}+\xi^2\left(\frac{1}{2}e^{-\lambda}\left[\frac{d\phi}{d\xi}\right]^2-\tilde{V}\right)\right],
\\
\label{eq_phi_ext}
&&\frac{d^2\phi}{d\xi^2}+\frac{2}{\xi}\left\{1+\frac{\sigma(n+1)e^{\lambda}}{2}
\left[\frac{2 v}{\xi}+\xi^2\left(\frac{1}{2}e^{-\lambda}\left[\frac{d\phi}{d\xi}\right]^2-\tilde{V}\right)-\frac{d v}{d\xi}\right]\right\}\frac{d\phi}{d\xi}
=e^{\lambda}\frac{d\tilde{V}}{d\phi},
\end{eqnarray}
where $e^{\lambda}$ is given by \eqref{lambda_dim}, and $\tilde{V}$ is taken from \eqref{poten_dim}.
Note that this system of equations \eqref{Einstein-00_cham_star_ext}-\eqref{sf_eq_ext} (or \eqref{eq_v_ext}-\eqref{eq_phi_ext})
is essentially the same
as those obtained in \cite{jetzer} which considered a real scalar field with quartic self-interaction.
The above system contains the parameter $\sigma$ as a trace
of the influence of the fluid on the external solution.
The solution sought beginning from the surface of the star
at $\xi=\xi_1$ using, as the boundary conditions,
the values of $v(\xi_1), \varphi(\xi_1)$ and $\nu(\xi_1)$
obtained from the solution of the equations
\eqref{eq_theta_app_n}-\eqref{eq_phi_dim_cham_star_n}
for the internal part of the configuration.
This allows one to determine the value of the integration constant $\nu_c$ from \eqref{nu_app}
by requiring $e^{\nu}$ to be equal to unity at infinity,
providing asymptotical flatness of the space-time.
(The values of $\nu_c$ for the examples shown in
figures \ref{metr_figs} are given in the caption.)
Thus the complete solution for the configuration under consideration is derived
by matching of the internal fluid solutions given by equations \eqref{eq_theta_app_n}-\eqref{eq_phi_dim_cham_star_n}
with the external solutions obtained from the system \eqref{eq_v_ext}-\eqref{eq_phi_ext}.

The system \eqref{eq_v_ext}-\eqref{eq_phi_ext} has obvious asymptotically flat solutions in the form:
$v \to \text{const} \approx v(\xi_1)$ with $v(\xi_1)$ taken from \eqref{lambda_dim_ext}
that corresponds to the fact that, by choosing the eigenvalues of the parameters $\Lambda$ and $\beta$
presented in tables \ref{tab1} and \ref{tab2}, the  external scalar field makes
negligible contribution to the total mass of the configuration. It allows using the solution for the metric functions in the form of
\eqref{lambda_dim_ext} as a good approximation; $\nu\to \text{const}$, where the constant is
made equal to zero by the corresponding choice of $\nu_c$ (see above);
$\phi \to \text{const}\, e^{-\mu \xi}/\xi$. The numerical calculations of the system
\eqref{eq_v_ext}-\eqref{eq_phi_ext} confirm this asymptotic behavior.

From figures  \ref{metr_figs}, \ref{energ_n_1_5_fig} and \ref{energ_n_1_0_fig} we can draw the following conclusions
about the metric functions, $e^{\lambda}, e^{\nu}$ and the mass distribution $M(\bar{\xi}/\bar{\xi_1})$:

(1) Asymptotically, as $\xi \to \infty$, the space-time becomes flat, i.e.
$e^{\nu}, e^{\lambda} \to 1$. To get this asymptotic behavior of the metric function
$e^{\nu}$, it is necessary to choose a specific
value for the central value, $\nu_c$ (see the caption of figures \ref{metr_figs}). The
asymptotic behavior of $e^{\lambda} \to 1$ follows from equation \eqref{lambda_dim} and the
fact that $v(\xi)|_{\xi\to 0} \to 0$ as $\xi ^3$.

(2) As shown in \cite{Tooper:1964}, the relativistic configurations without a scalar field are characterized by a greater concentration of matter
toward the center than in the non-relativistic case. By including a scalar field, we obtain even greater concentration of matter
toward the center. This can be seen by comparing figures \ref{metr_figs} with figures 1 and 2 from the paper \cite{Tooper:1964}.
The main reason for the increased concentration of mass is the non-minimal coupling  between the scalar field
and the  fluid with the choice of the coupling function given by \eqref{f_def}. The functional form of this coupling gives the required regular
solutions, lying in the range $0<\xi<\xi_1$, only for the eigenvalues of the parameter  $\beta \gg 1$
(see tables \ref{tab1}, \ref{tab2}). This leads to the fact that near the origin, when $\xi \approx 0$, a greater concentration
of mass occurs due to the presence of the term  $\beta\phi^2\theta^n$ in the expression for the energy density \eqref{tot_dens_poly_scalar}.
At the same time both $\phi$ and $\phi^\prime$ give relatively small contributions to the
energy density compared to the term coming from the non-minimal scalar-fluid coupling. This fact will become important, below,
when we consider the stability of the solutions.

(3) The total energy density presented in figures \ref{energ_n_1_5_fig} and \ref{energ_n_1_0_fig} strongly depends on the
polytropic index $n$. At comparable central values of the scalar field $\phi_0$  the energy density for the case
$n=1.5$ is several times greater than for the case $n=1.0$. In turn, both values of $n$ yield a greater energy density
at the center of the configurations than in the case of the relativistic stars without a scalar field
\cite{Tooper:1964} when the energy density was found to be 1.

(4) Despite the higher concentration of matter at the center of the configurations with the addition of the scalar field,
their masses are considerably smaller than the masses of the relativistic stars of the same size but without a scalar field
(see table \ref{tab2}). This occurs as a consequences of the fact that the external region of the stars, with only the
scalar field as a source, is strongly rarefied due to the rapid vanishing of the scalar field in the region $\xi \ge \xi_1$.

\subsection{Non-relativistic case}
\label{sec_non_rel_w_mass}

In this section we consider the non-relativistic limit of the system \eqref{eq_theta_app_n}-\eqref{eq_phi_dim_cham_star_n}.
The non-relativistic limit corresponds to $\sigma \to 0$, and $p \ll \rho$. First, we recall some results about this
system in the absence of the scalar field \cite{Zeld}. With no scalar field the system of equations
\eqref{eq_theta_app_n}-\eqref{eq_phi_dim_cham_star_n} reduces to the well known Lane-Emden equation
\begin{equation}
\label{Lane_Emden_eq}
\frac{1}{\xi^2}\frac{d}{d\xi}\left[\xi^2\frac{d\theta}{d\xi}\right]=-\theta^n ~.
\end{equation}
This equation has solutions which describe finite size configurations for different values of the parameter $n$ \cite{Zeld}.
The non-relativistic limit for the case when there a scalar field is obtained by omitting
all terms with $\sigma$ in equations \eqref{eq_theta_app_n}-\eqref{eq_phi_dim_cham_star_n} except for the term
$\beta \sigma \theta^{n+1}$ in the scalar field equation
\eqref{eq_phi_dim_cham_star_n}. This term must be kept since the product $\beta \sigma$ is generally a
non-zero quantity even for $\sigma \to 0$. Taking all the above into account we rewrite
the system \eqref{eq_theta_app_n}-\eqref{eq_phi_dim_cham_star_n} in the following form:
\begin{eqnarray}
\label{eq_theta_app_n_newt}
\xi^2\frac{d\theta}{d\xi}&=&
\xi^3\left[\frac{\beta}{2}\phi^2\theta^n+
 \mu^2 \phi^2+\frac{1}{2} \Lambda \phi^4-\frac{1}{\xi^2}\frac{d v}{d\xi}\right]-v
,\\
\label{eq_v_app_n_newt}
\frac{d v}{d\xi}&=&\frac{\xi^2}{2}\left\{\beta\phi^2\theta^n+
\left(\frac{d\phi}{d\xi}\right)^2+ \mu^2 \phi^2+\frac{1}{2} \Lambda \phi^4\right\},\\
\label{eq_phi_dim_cham_star_n_newt}
\frac{d^2 \phi}{d\xi^2}+\frac{2}{\xi}\frac{d\phi}{d\xi}&=&
\left(\mu^2-\beta \sigma\theta^{n+1}\right)\phi+\Lambda \phi^3.
\end{eqnarray}
Here, as in the relativistic case the presence of the term $\beta \sigma \theta^{n+1}$ is important --
it leads to a change in sign of the effective mass term $\left(\mu^2-\beta \sigma\theta^{n+1}\right)$ for
certain values of $\beta$. This feature is important for the existence of regular solutions.

Proceeding as in the previous section, we obtain numerical results for the non-relativistic limit
which we present in table \ref{tab3}. We chose the parameter $\sigma$ to be
$\sigma=0.001$. Such a small value of $\sigma$ requires that $\beta$ be large enough so that
the first term on the right hand side of equation \eqref{eq_phi_dim_cham_star_n_newt} is negative,
which is the necessary condition to have regular solutions. In the non-relativistic case, as in the relativistic case,
we need to find eigenvalues of two parameters $\beta$ and $\Lambda$ in order to have regular solutions for a
given values of the polytropic index, $n$, and the central value of the scalar field, $\phi_0$. From table \ref{tab3}
we do indeed find that there are values of $\beta$ large enough to make the first term on the
right hand side of \eqref{eq_phi_dim_cham_star_n_newt} negative thus yielding regular solutions.

\begin{table}[htbp]
 \caption{The parameters of the non-relativistic configurations with a variable central value of the scalar field
$\phi_0$. The observable radius of a star $\bar{\xi}_1$  is fixed. The choice of the parameters $\sigma$ and $n$
is indicated in the table.}
\begin{center}
\begin{tabular}{|p{2cm}|p{2cm}|p{2cm}|p{2cm}|p{2cm}|p{2cm}|p{2cm}|}
\hline
$\phi_0$&$\Lambda$ & $\beta$ & $\xi_1$ & $v(\xi_1)$&$\rho_c/\bar{\rho}$&$-\Omega/M$\\
\hline
\multicolumn{7}{|c|}{$\sigma=0.001, n=1.5$,  $\bar{\xi}_1=3.6560$}\\
\hline
\multicolumn{7}{|c|}{\bf Non-relativistic star without a scalar field }\\
\hline
&   &    &    3.6538&	2.7141&	5.9907&		0.0000\\
\hline
\multicolumn{7}{|c|}{\bf Non-relativistic star with the scalar field}\\
\hline
0.039	&1040	&20600	&3.6538	&0.4384	&37.0928	&-0.0009\\
0.038	&1850	&21800	&3.6538	&0.4373	&37.1840	&-0.0009\\
0.036	&3980	&24500	&3.6538	&0.4354	&37.3411	&-0.0009\\
0.034	&7010	&27600	&3.6538	&0.4346	&37.4105	&-0.0009\\
0.030	&17575	&35500	&3.6538	&0.4349	&37.3895	&-0.0009\\
0.028	&25895	&39500	&3.6538	&0.4462	&36.4388	&-0.0009\\
0.027	&31270	&41700	&3.6538	&0.4556	&35.6855	&-0.0009\\

\hline
\multicolumn{7}{|c|}{$\sigma=0.001, n=1.0$,  $\bar{\xi}_1=3.1437$}\\
\hline
\multicolumn{7}{|c|}{\bf Non-relativistic star without a scalar field }\\
\hline
&   &    &    3.1416&	3.1416&	3.2899&		0.0000\\
\hline
\multicolumn{7}{|c|}{\bf Non-relativistic star with the scalar field}\\
\hline
0.039	&190	&10700	&3.1416	&0.6568	&15.7364	&-0.0008\\
0.038	&595	&11000	&3.1416	&0.6687	&15.4556	&-0.0008\\
0.036	&1645	&12000	&3.1416	&0.6761	&15.2872	&-0.0008\\
0.034	&3070	&13100	&3.1416	&0.6868	&15.0493	&-0.0008\\
0.030	&7645	&15700	&3.1416	&0.7166	&14.4227	&-0.0008\\
0.025	&20685	&21000	&3.1416	&0.7457	&13.8609	&-0.0008\\
\hline
\end{tabular}
\end{center}
\label{tab3}
\end{table}

\begin{figure}[t]
\begin{minipage}[t]{.51\linewidth}
  \begin{center}
  \includegraphics[width=9cm]{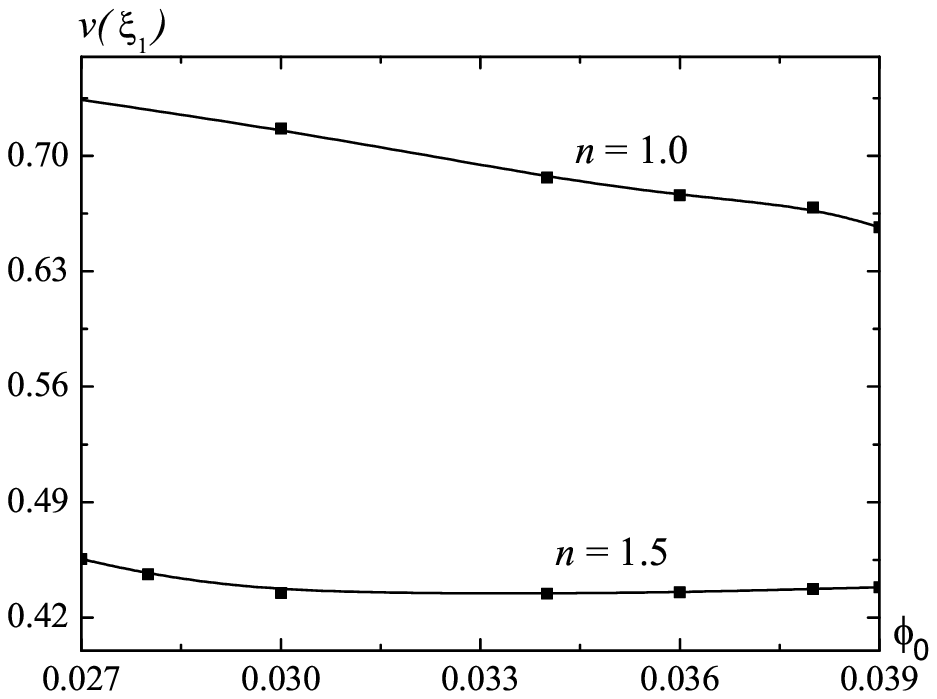}
  \caption{\small The dependence of the function $v(\xi_1)$
  on the   central value of  $\phi_0$ at the fixed  $\bar{\xi}_1=3.6560$ for $n=1.5$
  and $\bar{\xi}_1=3.1437$ for $n=1.0$, and $\sigma=0.001$ for both graphs.
The data are taken from table \ref{tab3}.}
    \label{mass_phi_fig_non_rel}
  \end{center}
\end{minipage}\hfill
\begin{minipage}[t]{.47\linewidth}
  \begin{center}
  \includegraphics[width=6.5cm]{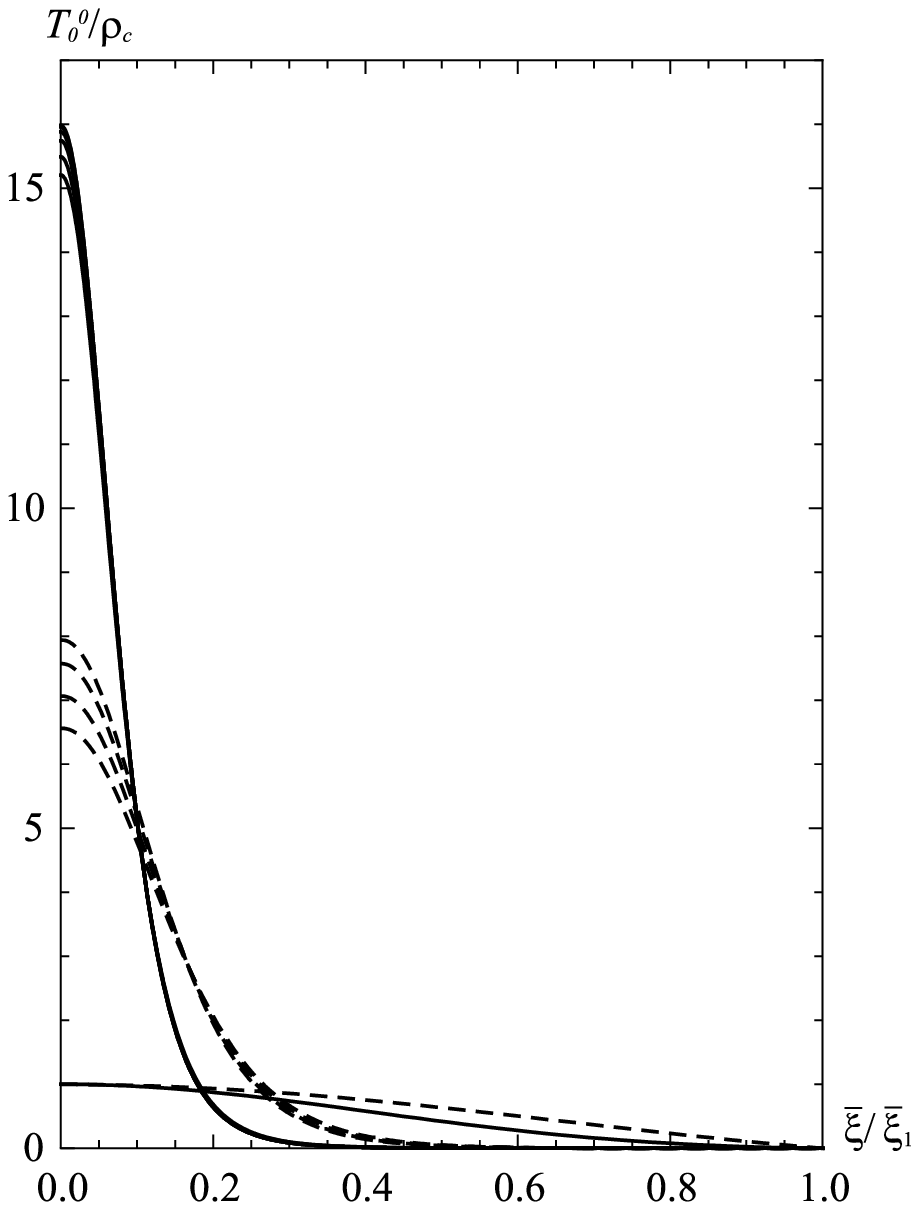}
  \caption{\small The plot of the total energy density, $T^0_0$, in units of $\rho_c$ in the non-relativistic case ($\sigma=0.001$)
  for different central values of $\phi_0$ from table \ref{tab3}. The solid lines correspond to the case of $n=1.5$, the dashed lines
  to the case of $n=1.0$. The top two sets of curves correspond to the configurations with the scalar field.
  For comparison the energy densities of the non-relativistic configurations without a scalar field $T_0^0=\rho_c \theta^n$
  are also shown by the two bottom, less peaked curves.}
 \label{energ_non_rel_fig}
  \end{center}
\end{minipage}\hfill
\end{figure}

Using the data from table \ref{tab3}, the dependence of the $v(\xi_1)$ on the central value of
the scalar field $\phi_0$ is given in figure \ref{mass_phi_fig_non_rel}. Figure \ref{energ_non_rel_fig}
uses the date from table \ref{tab3} to plot the ratio of the energy density to the central
density, $T^0_0 / \rho _c$ as a function of the dimensionless, normalized radius.
Also from table \ref{tab3} one can see that we have fixed the radius of a star $R=\xi_1$ to be equal
to the size of a non-relativistic star without a scalar field. From the figure we
can draw the following observations:

(1) The masses of the stars with a scalar field are considerably smaller than the masses
of the stars without a scalar field. In the case $n=1.0$ the masses are
4-5 times smaller, and in the case $n=1.5$ about 6 times smaller.

(2) The mass is a fairly slowly varying function of $\phi_0$, especially when $n=1.5$.

(3) There is a greater concentration of the mass density towards the center of the stellar configurations.
This can be seen explicitly by comparing with the distribution of the mass density for a star without a scalar field which
is also presented in figure \ref{energ_non_rel_fig}.

All these results are very similar to those of the relativistic case.

\subsection{Stability of the solutions}
\label{stab_sec}

In this section we discuss the issue of the stability of the regular solutions obtained above.
There are two basic approaches to studying the stability: (i)
The energy approach to the theory of equilibrium for a star \cite{Zeld};
(ii) A more rigorous dynamical stability approach based on studying the stability of linear and nonlinear
time-dependent perturbations. In this paper we will use the first approach.  Proceeding along the lines of reference
\cite{Tooper:1964}, we define the total energy $E$ of the system, including the internal and gravitational
energies, as
\begin{equation}
\label{total_energ}
E=M=4\pi \int_0^R T_0^0 r^2 dr,
\end{equation}
where $T_0^0$ is defined by \eqref{emt_cham_star} and corresponds to the total energy density of the system
(recall that $c=1$). Next, we consider a system consisting of a gas of particles
having a rest-mass density $\rho_g$. Its
relativistic energy density is the sum of the rest energy $\rho_g$ plus the density of internal energy.
For the special case of an adiabatic process which assumes the absence of heat flow terms in the energy-momentum
tensor \eqref{emt_cham_star},
it is possible to obtain a relation between the gas density $\rho_g$
and the total mass density $\rho_t \equiv T_0^0$
or alternatively between $\rho_g$ and  $\theta $, $\varphi$.

To do this we use the first law of thermodynamics which, in our case, takes the form
\begin{equation}
\label{first_law}
d\rho_t+(\rho_t+p_t)\frac{d V}{V}=0,
\end{equation}
where $p_t \equiv -T_1^1$ is the total pressure,  and $V$ is the specific volume
(this symbol $V$ should not be confused with the potential energy used earlier).
Since $dV/V=-d\rho_g/\rho_g$,
equation \eqref{first_law} gives the following relation
\begin{equation}
\label{first_law_2}
\frac{d\rho_g}{\rho_g}=\frac{d\rho_t}{\rho_t+p_t}.
\end{equation}
In general, when $\rho_t$ and $p_t$ are functions of $\theta, \varphi$, this equation cannot be integrated.
However, for our configuration, the numerical calculations performed in the previous sections show that the main part of the energy
is provided by the term containing the non-minimal coupling but not by
$\varphi^\prime$ and $V(\varphi)$. This allows us to neglect the terms containing the scalar field kinetic energy
and potential as compared with $f \rho$ and $f p$ in the $T_0^0$ and $T_1^1$ components of the energy-momentum tensor \eqref{emt_cham_star},
respectively. In this approximation we have the following expressions for the total
mass density and the pressure:
$$
\rho_t \approx f \rho, \quad p_t \approx f p.
$$
Substituting these expressions into \eqref{first_law_2} and taking into account that
$\rho=\rho_c \theta^n$ and $p=\sigma \rho_c \theta^{n+1}$, we have from \eqref{first_law_2}:
$$
\frac{d\rho_g}{\rho_g}=\frac{n d\theta}{\theta(1+\sigma \theta)}+\frac{d f}{f(1+\sigma \theta)}.
$$
This equation differs from the case without a scalar field by the presence of the second term on the right-hand side
containing the  coupling function $f$. In the absence of the non-minimal coupling, i.e. when $f=1$, this term vanishes.
Integrating, we find
\begin{equation}
\label{rest_mass_den}
\rho_g=\rho_{gc}\left[\frac{(1+\sigma)\theta}{1+\sigma\theta}\right]^n \exp{\left[\int \frac{d f}{f(1+\sigma \theta)}\right]},
\end{equation}
where $\rho_{gc}$ is the value of the gas density $\rho_g$ at the center of the configuration. This constant may be evaluated as
follows \cite{Tooper:1964}: near the boundary of the configuration, where $\theta \ll 1$, equation \eqref{rest_mass_den} becomes
approximately
$$
\rho_g \approx \rho_{gc}(1+\sigma)^n \theta^n \exp{\left(\int_{\xi_a}^\xi \frac{d f}{f}\right)},
$$
where $\xi_a$ is the point starting from which the approximation $\theta \ll 1$ becomes valid.
This equation can be rewritten in differential form as
\begin{equation}
\label{rest_mass_diff}
\frac{d}{d\xi}\left\{\ln{\left[\frac{\rho_g}{\rho_{gc}(1+\sigma)^n\theta^n}\right]}\right\}\approx
\frac{d}{d\xi}\left(\int_{\xi_a}^\xi\frac{d f}{f}\right).
\end{equation}
Since $f=f(\phi(\xi))$ is a function of $\xi$,
the term on the right-hand side can be evaluated as follows
$$
\frac{d}{d\xi}\left(\int_{\xi_a}^\xi\frac{d f}{f}\right)\equiv
\frac{d}{d\xi}\left(\int_{\xi_a}^\xi\frac{(d f/d\xi) d\xi}{f}\right)=\frac{d}{d\xi}\left(\ln{f}\right).
$$
Taking this expression into account, equation \eqref{rest_mass_diff} takes the form
$$
\rho_g \approx \rho_{gc}(1+\sigma)^n f \theta^n.
$$
Near the boundary the internal energy density is small compared to the rest-mass energy density, so that $\rho_t \approx \rho_g$.
Comparing the above equation with the expression $\rho_t \approx f \rho_c \theta^n$, we obtain
$$
\rho_{gc}=\frac{\rho_c}{(1+\sigma)^n},
$$
and equation \eqref{rest_mass_den} becomes
\begin{equation}
\label{rest_mass_den_2}
\rho_g=\rho_c\left[\frac{\theta}{1+\sigma\theta}\right]^n \exp{\left[\int \frac{d f}{f(1+\sigma \theta)}\right]}.
\end{equation}

Using  expression \eqref{rest_mass_den_2}, the corresponding proper energy $E_{0g}$ of the gas may be defined as the integral of $\rho_g$
with respect to proper volume $dV=4\pi r^2 e^{\lambda/2}dr$ for the metric \eqref{metric_sphera} as follows
\begin{equation}
\label{proper_energ_gas}
E_{0g}=M_{0g}=4\pi \int_0^R \rho_g e^{\lambda/2} r^2 dr.
\end{equation}
The quantity $M_{0g}$ is proportional to the total number of particles $N$ in the configuration, $M_{0g}\equiv N/A$,
where $A$ is Avogadro's number. Using the dimensionless variables \eqref{dimless_xi_v} and expressions \eqref{lambda_dim},
\eqref{rest_mass_den_2}, we obtain the proper energy of the gas \eqref{proper_energ_gas} in units of the total energy $E=M$
in the following form
\begin{equation}
\label{proper_energ_gas_2}
\frac{E_{0g}}{E}=\frac{1}{v(\xi_1)}\int_0^{\xi_1}
\frac{\theta^n \exp{\left\{\int_0^{\xi_1}\left[f(1+\sigma\theta)\right]^{-1}df\right\}}\xi^2 d\xi}{\left[1+\sigma\theta\right]^n
\left[1-2\sigma(n+1)v/\xi\right]^{1/2}}.
\end{equation}
This expression differs from the corresponding expression in \cite{Tooper:1964} through the presence of an extra factor
$\exp\{...\}$ under the integral which comes from
the non-minimal coupling in the system. Stability of the configuration can tested via the sign
of the expression \cite{Tooper:1964,Zeld}
\begin{equation}
\label{bind_energ}
\frac{\text{Binding~Energy}}{E}=\frac{E_{0g}-E}{E} =\frac{E_{0g}}{E}-1.
\end{equation}
The necessary (but not a sufficient) condition for stability of the system that $E_{0g} > E$ i.e. that
the total system energy $E$ is less that the energy of the non-gravitationally interacting gas, $E_{0g}$, making
$E$ the energetically preferred state. The condition in \eqref{bind_energ} amounts to requiring
that the Binding Energy defined in this equation be positive. In the paper \cite{Tooper2}, the question of stability
of adiabatic polytropic configurations was considered. It was shown that such configurations may have both positive and
negative binding energies depending on the value of the polytropic index $n$, and the parameter
$\sigma$ (configurations with  $1\leq n \leq 3$ and $0.01 \leq \sigma \leq 100$ were considered).

When a non-minimally coupled scalar field is included, we considered a very narrow range of these parameters, $\sigma=0.2$ and $n=1.0,\, 1.5$,
and also only one choice of the coupling function $f$ given by the form \eqref{f_def}. In this case the numerical calculations indicate that
$E_{0g}/E \ll 1$, i.e. we have the negative binding energy, and correspondingly the configurations under consideration are unstable.
The same values of $\sigma$ and $n$ without a scalar field give configurations with the positive binding energy  \cite{Tooper2}.
Obviously, the difference between the present results and those of \cite{Tooper2} is connected with the presence of the extra factor $\exp\{...\}$ under the integral
in the expression \eqref{proper_energ_gas_2}  whose value, as numerical calculations indicate,
is much less than unity for the parameters  $\sigma$, $n$ used in the paper, and for the coupling function $f$ of the form \eqref{f_def}.

The behavior of $\exp\{...\}$ in \eqref{proper_energ_gas_2} can be approximately estimated as follows:
Since we are looking only for solutions with $0\leq \theta \leq 1$,  the value of
$(1+\sigma \theta)$ may be approximated as giving a finite contribution to the value of integral
 $\int_0^{\xi_1}\left[f(1+\sigma\theta)\right]^{-1}df$ in the form of some constant factor
 $\alpha>0$, i.e.  $\int_0^{\xi_1}\left[f(1+\sigma\theta)\right]^{-1}df \approx \alpha \int_0^{\xi_1} d \ln{f}$.
Then one can see that
$$
\exp{\left\{\int_0^{\xi_1}\left[f(1+\sigma\theta)\right]^{-1}df\right\}} \approx \left[\frac{f(\xi_1)}{f(0)}\right]^\alpha.
$$
When the coupling function $f$ is taken in the form \eqref{f_def}, we have the finite value of
$f(0)$, and since $\phi(\xi_1) \to 0$ then $f(\xi_1) \to 0$ as well, thereby suppressing the value of
$E_{0g}/E$ in \eqref{proper_energ_gas_2}. In the non-relativistic case, where $\sigma \to 0$, we have $\alpha \to 1$ and
the situation with the instability is the same.

Possible ways of resolving this problem are:

(i) Consider configurations for which
$\phi(\xi_1)\neq 0$, and the function  $f(\xi)$ varies slowly in the range $0<\xi<\xi_1$ to provide a large value
of the above $\exp\{...\}$. In this case, one can expect that the configuration will have a ``tail'' of a scalar field
outside the fluid at $\xi>\xi_1$ (a similar thing happens for configurations considered in
\cite{Dzhunushaliev:2011xx}). In this case, it will be necessary to perform a stability analysis of both
internal solutions (where the fluid and the scalar field are involved) and external solutions (where only the scalar field is involved).

(ii) Since the solutions and their stability are sensitive to the nature of the scalar field-fluid coupling, which in this 
paper was take to have the form given in \eqref{f_def}, one could look for stable solutions by examining different forms of
this coupling; for example one could try a non-polynomial coupling of the form $f=e^{\phi}$.
These stability studies will be the focus of future work.

\section{Non-relativistic case: an analytical solution for a massless scalar field}
\label{non_rel_analyt}

The numerical results obtained in the previous sections demonstrate  the possibility of obtaining the regular solutions
both in relativistic and non-relativistic cases when the coupling function $f$ is chosen in the form of \eqref{f_def}.
This choice is not the only one possible. It will be shown latter in this section that, by choosing the special form
of the function $f$, it is possible to find an analytical solution in a particular non-relativistic case when
the polytropic index $n=0$, and the scalar field is taken to be massless. This case corresponds to an incompressible fluid
with a constant mass density, $\rho=\rho_c=const$, and a spatial varying pressure, $p \neq const$.

To begin we rewrite the system of equations
\eqref{eq_theta_app_n}-\eqref{eq_phi_dim_cham_star_n} for the massless case and
in the non-relativistic limit ($\sigma \to 0$) in the form
\begin{eqnarray}
\label{eq_theta_app_n_newt_massless}
\xi^2\frac{d\theta}{d\xi}&=&
-\frac{\xi^3}{2} \left( \frac{d \phi}{d \xi} \right)^2 -v,\\
\label{eq_v_app_n_newt_massless}
\frac{d v}{d\xi}&=&
\xi^2\left[
	f \theta^n+ \frac{1}{2}\left( \frac{d \phi}{d \xi} \right)^2
\right],\\
\label{eq_phi_dim_cham_star_n_newt_massless}
\frac{d^2 \phi}{d\xi^2}+\frac{2}{\xi}\frac{d\phi}{d\xi}&=&
-\sigma \theta^{n+1}\frac{d f}{d\phi}.
\end{eqnarray}
We have kept the $\sigma$-containing-term on the right-hand side of equation \eqref{eq_phi_dim_cham_star_n_newt_massless}
since it is does not necessarily small for {\it arbitrary}  $f$.
Differentiating the first equation of the system, and substituting the second and the third equations
into this expression, gives
\begin{equation}
\label{theta_non_rel_massless}
\frac{1}{\xi^2}\frac{d}{d\xi}\left(\xi^2\frac{d\theta}{d\xi}\right)=
-f \theta^n+\sigma \xi \theta^{n+1}\frac{d f}{d\xi}.
\end{equation}
The last term of this equation contains the factor  $df/d\xi \equiv \phi^{\prime} df/d\phi$. For $f=1$,
equation \eqref{theta_non_rel_massless} reduces to the Lane-Emden equation
\eqref{Lane_Emden_eq} which has an analytical solution for the case of an incompressible fluid with
a constant  mass density $\rho=\rho_c=const, p \neq const, n=0$ -- see \cite{Zeld}, equation (10.3.14).
This solution is
\begin{equation}
\label{sol_non_rel_class}
\theta=1-\frac{1}{6}\xi^2.
\end{equation}
Equation \eqref{theta_non_rel_massless} can also be integrated analytically by choosing the function
$f$ to have the following power-law form
\begin{equation}
\label{f_non_rel_power}
f=f_0+\frac{\beta}{m}\xi^m,
\end{equation}
where $f_0$, $\beta$ and $m$ are arbitrary parameters. Using this function and taking the
parameter $m$ to be of order 1 so that the product $\sigma \theta \ll |1/m|$, we have from
\eqref{theta_non_rel_massless}:
$$
\frac{1}{\xi^2}\frac{d}{d\xi}\left[\xi^2\frac{d\theta}{d\xi}\right]=-\left(f_0+\frac{\beta}{m}\xi^m\right)\theta^n.
$$
When $n=0$, this equation can be integrated giving the following regular solution
\begin{equation}
\label{sol_non_rel_scalar}
\theta=\theta_0-\frac{f_0}{6}\xi^2-\frac{\beta}{m(m+2)(m+3)}\xi^{m+2}.
\end{equation}
Here $\theta_0$ is an integration constant. A second integration constant was set equal to zero to make the solution
regular at $\xi=0$. We take $\theta_0=1 $ so that in the absence of the scalar field (i.e. when $f_0 =1$ and $\beta =0$) equation
\eqref{sol_non_rel_scalar} will reduce to equation \eqref{sol_non_rel_class}.
This solution is regular everywhere, including the point $\xi=0$,
when  $m>-2$. The degenerate case $m=0$ corresponds to the classical solution
\eqref{sol_non_rel_class} which can be obtained from \eqref{sol_non_rel_scalar} by setting
$f_0=1$ and $\beta=0$. This choice  that corresponds to turning equation
\eqref{theta_non_rel_massless} into the Lane-Emden equation \eqref{Lane_Emden_eq}.

We now analyze the behavior of the solution as  $\xi \to 0$. To begin we calculate the first and second
derivatives of $\theta$:
$$
\frac{d\theta}{d\xi}=-\frac{f_0}{3}\xi-\frac{\beta}{m(m+3)}\xi^{m+1}, \quad
\frac{d^2\theta}{d\xi^2}=-\frac{f_0}{3}-\frac{\beta(m+1)}{m(m+3)}\xi^{m}.
$$
For a spherically symmetric solution with maximal mass density as $\xi \to 0$, two conditions must be satisfied:
(1) the first derivative must be equal to zero; (2) the second derivative must be negative. The former condition
gives $m>-1$. Given this the behavior of the second derivative is:

\begin{tabular}{ll}
if $(-1<m<0)$: \; $\theta^{\prime\prime}\to$  &  $\left\{  \begin{tabular}{l}
$+\infty$ \; for $\beta > 0$;\\[\medskipamount]
$-\infty$ \; for $\beta < 0$;\\
\end{tabular}  \right.  $
\\[\bigskipamount] \\
if $m>0$: \qquad \qquad  $\theta^{\prime\prime}\to$&\,\,\, $-f_0/3$ \;for any $\beta$.\\[\medskipamount]
\end{tabular}

\noindent One can see that for $(-1<m<0)$ the second condition is satisfied only if $\beta <0$.

On the other hand, the regularity of the solution assumes the presence of a point on the axis,
$\xi$, where $\theta=0$. For the classical solution
\eqref{sol_non_rel_class}, this point is $\xi_1|_{\theta=0}=\sqrt{6}$.
In the case of configurations with a scalar field, the location of the point $\xi_1|_{\theta=0}$
will depend on the factors in front of the $\xi^{m+2}$ term in \eqref{sol_non_rel_scalar}, i.e. it will be
determined by the values of $\beta$ and the denominator  $m(m+2)(m+3)$.
In the range $(-1<m<0)$ only negative $\beta$ are allowed if one wants $\theta '' <0$ as $\xi \to 0$,
which is one of conditions that the solution \eqref{sol_non_rel_scalar} be regular.
In the range $m>0$ the regularity of the solution is guaranteed if
$\beta>0$, since otherwise  the solution grow as a power law of $\xi$.

The above analysis shows that the inclusion of a scalar field decreases the size of configurations,
as compared to the classical case \eqref{sol_non_rel_class}, for all acceptable values of the parameter $m$.
In this sense the behavior of the solutions with a massless scalar field and
the coupling function in the form of \eqref{f_non_rel_power} differs
from the behavior of the solutions obtained in section \ref{sec_non_rel_w_mass}
when the function $f$ was chosen to have the form \eqref{f_def}.
In the latter case, the size of the configurations for given values of $n$ and $\sigma$
depends considerably on the parameters $\Lambda$,  $\beta$, and $\phi_0$, and it can be larger or
smaller than the size of a configuration without a scalar field.

Substituting the expression for $\theta$ from \eqref{sol_non_rel_scalar} into the scalar field equation
\eqref{eq_phi_dim_cham_star_n_newt_massless} with $n=0$, we find the following analytical solution for the
scalar field $\phi$ in the interior of the configuration
\begin{equation}
\label{sol_phi_non_rel_scalar}
\phi _{in}=\phi_0+\frac{\beta\sigma }{6}\left[
\frac{3\beta \xi^{2+m}}{m(m+2)^2(m+3)(2m+3)}+\frac{f_0 \xi^2}{(m+3)(m+4)}-\frac{6}{(m+1)(m+2)}
\right]\xi^{m+1}.
\end{equation}
One can see from this expression that the behavior of the scalar field is defined in large measure by
the factor $\beta\sigma$ in front of the square brackets.
Since in the non-relativistic limit $\sigma \to 0$, this factor will differ appreciably from zero only for large $\beta$.
For $\beta$ not large one can expect that the field will be practically constant and equal to the central value
$\phi_0$ up to the boundary of the fluid at $\xi=\xi_1$ where $\theta=0$. The general solution for $\phi$ for all
$\xi$ is obtained by matching of the internal solution \eqref{sol_phi_non_rel_scalar} with an external solution of
equation \eqref{eq_phi_dim_cham_star_n_newt_massless} where the right-hand side is equal to zero.
The exterior equation has a solution of the form
$$\phi_{ext}=C_1+C_2/\xi,$$
where $C_1, C_2$ are integration constants which are determined by matching of the external solution
$\phi_{ext}$ with the interior solution $\phi _{in}$ solution at the boundary of the fluid at $\xi=\xi_1$.
One can see from this expression that the corresponding mass density, which is proportional to $\phi_{ext}^{\prime 2}$,
tends asymptotically to zero.

\section{Conclusion}
\label{sec_concl}

In this article we studied gravitating, spherically symmetric, star-like, configurations with a matter source
consisting of a normal (i.e. non-ghost, non-phantom) scalar field  plus
a  perfect isotropic fluid. The motivation for studying such a model is that
scalar fields have found broad use in various cosmological models as well as astrophysical models.
It is natural to postulate that, if scalar fields do really play a role in the Universe
that these scalar fields might play a role in the structure of compact objects such as Main Sequence stars or
neutron stars.  Proceeding from this assumption, we studied a model of a star-like
configuration supported by a scalar field non-minimally coupled to ordinary matter in the form of a perfect fluid.
As an example, we considered the case when the scalar field had a quadratic mass term and a quartic self-interaction
giving a scalar field potential of the form \eqref{poten_dim}. The coupling between the scalar field and the perfect fluid, $f$,
was taken to have the form \eqref{f_def}. For this model we studied solutions both in the relativistic and non-relativistic limit.
Our investigation showed that the existence of regular solutions is possible exactly because of the presence of the non-minimal coupling
between the scalar field and fluid. In the absence of such coupling, the potential \eqref{poten_dim} gives only
singular solutions \cite{jetzer}. From the mathematical point of view, the existence of regular solutions in our model is possible
because of the appearance in  scalar field equation \eqref{eq_phi_dim_cham_star_n} of an effective mass term
$m_{eff}=\left(\mu^2-\beta \sigma\theta^{n+1}\right)$ whose sign depends both on  the behavior of the fluid density
$\theta$ and the values of the parameters $\beta$ and $\sigma$.
It was shown that when $m_{eff}<0$, there were regular solutions with finite masses and sizes.
In some sense the solutions presented above are a cross between the interacting scalar field solutions of
\cite{Colpi} and the interacting real scalar field solutions of \cite{jetzer}. In the interior region, $0<\xi <\xi_1$,
we have a real scalar field and a fluid which interacts with the scalar field. Due to the chameleon-like behavior of
the scalar field with respect to the fluid (from figure \ref{metr_figs} one can see that in the interior region the field mimics the fluid)
one might think of the two degrees of freedom associated with the scalar field and the fluid as being equivalent to
the two degrees of freedom of a complex scalar field. While there is certainly some validity in this analogy (both the
present solutions and those in \cite {Colpi} are non-singular) one must avoid pushing the analogy too far since the
fluid vanishes exactly at some point (i.e. $\theta =0$ at $\xi = \xi_1$) while the complex scalar field of
\cite{Colpi} and the real scalar of the present solution only go to zero asymptotically. Thus after reaching the point
$\xi =\xi _1$ our solutions become those of \cite{jetzer}.

Our results are interesting since they show that by adding such a fluid one can
find regular solutions, thus evading some gravitational version of Derrick's theorem \cite{Bronnikov:2002kw,Bronnikov:2005gm}
which prohibits regular solutions for the system of gravity plus a normal scalar field if
the potential $V(\varphi)>0$. Thus the addition of the fluid was crucial to the existence of these solutions.
The original, non-gravitational version of Derrick's theorem \cite{derrick,rajaraman}
assumes not merely the regularity of solutions, but also their stability.
In this paper we have performed a preliminary stability analysis
based on energy considerations. In this case we compared the total energy of the system
(including the internal and gravitational energies) and the rest energy of the gas particles.
For the values  of the polytropic index $n$ and the parameter $\sigma$, and the specific coupling function $f$
in the form of \eqref{f_def}, this energy approach showed that the solutions studied in this paper are unstable. This was because
the binding energy of the system, which was equal to the difference of the rest energy and the total energy,
was negative. But, as it was shown in \cite{Tooper2}, even a positive binding energy does not guarantee stability
of a system allowing its transition to an energetically  more advantageous state with the same
polytropic index $n$, but having another parameter $\sigma$.
Obviously, such process is accompanied  by  an ejection of the excess of energy. In our case, the possibility is not excluded
that there exist regular solutions with positive binding energy, for other values
of $n$ and $\sigma$ and other parameters of the scalar field
$\beta, \Lambda, \phi_0$. There is an additional possibility for finding stable solutions which was described at the end
of section \ref{stab_sec} -- since the solutions and their stability are sensitive to the form of the scalar field-fluid 
coupling (which in the present work had the form \eqref{f_def}) one could look for stable solutions by changing the
form of the scalar field-fluid coupling. For example, one could try non-polynomial couplings of the form
$f(\phi) = e^{\phi}$. Finally, the energy approach to stability which was used in this paper,
can and should be supplemented  by a study of dynamical stability along the lines
of the linear stability analysis performed in \cite{jetzer, clayton, Dzhunushaliev:2010bv}
or by using the catastrophe theory method suggested in \cite{Kusmartsev:1991pm}. In future
work we plan to perform a stability analysis of the chameleon star solutions investigated here 
using both these approaches.

The spherically symmetric solution that we found for the system considered in this paper were called chameleon stars
in analogy with chameleon cosmological models \cite{Farajollahi:2010pk} because the characteristics of the scalar field
(e.g. its mass) strongly depend the other fields and fluids in its environment. We characterized
the behavior of these chameleon star configurations for both relativistic and non-relativistic cases. For
the relativistic chameleon stars we found that in general the mass-energy density tended to be more concentrated
toward the center of the star at $\xi =0$. Despite this greater concentration of mass near the center of the
chameleon star the total mass of the chameleon star was lower than the corresponding relativistic stars without
a scalar field. This was the result of the outer regions of the chameleon star being much less dense than a corresponding
non-chameleon, relativistic star. Similar comments apply to the non-relativistic chameleon stars. For the
non-relativistic case we were able to find an analytical solution for the case when the scalar field
was massless. This analytical solution had similar behavior to the numerically obtained solutions
from section \ref{sec_non_rel_w_mass}.

We briefly give some speculations about possible physical applications of the present solutions. Scalar fields are thought to play
a role in cosmological dynamics (dark energy) and in dynamics at the galactic scale (dark matter). The present proposal
is that a scalar field could play some role in astrophysics at the stellar scale in the formation of the chameleon star
configurations discussed here. The fluid in our model would be provided by the star. For example one could consider
neutron stars which contain a significant amount of the scalar field. From the discussion above our chameleon stars --
both relativistic and non-relativistic -- would be less massive for a given radius despite having a higher concentration of mass
near the center. Thus one might look for neutron stars which have a larger radius for a given mass than would be expected for a
normal neutron star. In this sense chameleon stars would have the opposite behavior from hypothetical quark stars which
have a larger mass for a given radius. Another possibility is that some living stars (i.e. stars which are still fusing elements
and are on the Main Sequence) might have trapped some significant amount of scalar field in their interior. Such living, chameleon stars
would tend to have a larger density near their center thus increasing the rate at which they fused elements. The consequence of this
is that such stars would be hotter and live a shorter than expected life span for their mass. Again a given chameleon star would
have less overall mass than a non-chameleon star but would be nevertheless have a higher, interior temperature.
The physical applications of the chameleon star model suggested above would of course require the existence of stable solutions.
A search for such stable solutions is a goal of future studies.

\section*{Acknowledgements}
V.D. and V.F. are grateful to the Research Group Linkage
Programme of the Alexander von Humboldt Foundation for the
support of this research.

\appendix
\numberwithin{equation}{section}

\boldmath
\section{Derivation of the energy-momentum tensor}
\unboldmath
\label{appen_emt}

In the case of the isentropic quasipotential flow, the Lagrangian of the continuous medium  has the form
 \cite{Stanuk1964,Stanuk}
\begin{equation}
\label{lagr_matter}
L_m=p={\cal W}-\varepsilon,
\end{equation}
where ${\cal W} =(\varepsilon+p)$ is the heat function, and $\varepsilon$ is the energy density. Introducing the quasipotential
$$
c S_i=c\frac{\partial S}{\partial x^i}={\cal W} u_i,
$$
where $c$ is the velocity of light and $S$ is the action for the matter, we have
$$
S_i S^i=\frac{w^2}{c^2}u_i u^i=\frac{{\cal W}^2}{c^2},
$$
from which it  follows
$$
{\cal W=}c\sqrt{g^{ik}S_i S_k}.
$$
The Lagrangian \eqref{lagr_matter} now takes the form
\begin{equation}
\label{lagr_matter_n}
L_m=c \sqrt{g^{ik}S_i S_k}-\varepsilon.
\end{equation}
The total action (including gravity plus all matter sources) in curvilinear coordinates is \cite{Land1}
\begin{equation}
\label{action_gen}
S=\frac{1}{c}\int L \sqrt{-g}d^4 x,
\end{equation}
where the Lagrange density, $L$, above is that given in equation \eqref{lagran_cham_star}. By varying the action
\eqref{action_gen} with respect to the metric, $g_{ik}$, one can obtain the Einstein equations and the energy-momentum tensor.
In this way the energy-momentum tensor is given by the expression
\begin{equation}
\label{emt_gen}
\frac{1}{2}\sqrt{-g} T_{ik}=\frac{\partial \sqrt{-g}\mathfrak{L}}{\partial g^{ik}}-
\frac{\partial}{\partial x^l}\frac{\partial \sqrt{-g}\mathfrak{L}}{\partial \frac{\partial g^{ik}}{\partial x^l}},
\end{equation}
where the Lagrangian, $\mathfrak{L}$, contains only the matter components of the Lagrangian \eqref{lagran_cham_star} namely
$$
\mathfrak{L}=\frac{1}{2}g^{ik}\partial_{i}\varphi\partial_{k}\varphi -V(\varphi)+f(\varphi) L_m.
$$
Since $\sqrt{-g}\mathfrak{L}$ does not depend on $\frac{\partial g^{ik}}{\partial x^l}$, the last term
in \eqref{emt_gen} equals zero. Then we have
\begin{equation}
\label{emt_gen_1}
\frac{1}{2}\sqrt{-g} T_{ik}=\frac{\partial \sqrt{-g}\mathfrak{L}}{\partial g^{ik}}=
\mathfrak{L}\frac{\partial \sqrt{-g}}{\partial g^{ik}}+
\sqrt{-g}\frac{\partial \mathfrak{L}}{\partial g^{ik}}.
\end{equation}
Taking into account that $dg=g g^{ik}dg_{ik}=-g g_{ik}dg^{ik}$ and also using the expressions \eqref{lagr_matter} and
\eqref{lagr_matter_n}, we find
$$
\mathfrak{L}\frac{\partial \sqrt{-g}}{\partial g^{ik}}=-\frac{1}{2}\sqrt{-g}g_{ik}\mathfrak{L},
$$
and
$$
\sqrt{-g}\frac{\partial \mathfrak{L}}{\partial g^{ik}}=\sqrt{-g}\left(\frac{1}{2}\frac{c S_i S_k}{\sqrt{g^{ik}S_i S_k}}f+
\frac{1}{2}\partial_i \varphi \partial_k \varphi\right)=
\frac{1}{2}\sqrt{-g}\left(f {\cal W} u_i u_k+
\partial_i \varphi \partial_k \varphi\right).
$$
Substituting these expressions into \eqref{emt_gen_1}, we finally have:
\begin{equation}
\label{emt_gen_fin}
T_{ik}=f\left[(\varepsilon+p)u_i u_k-g_{ik}p\right]+\partial_i \varphi \partial_k \varphi-
g_{ik}\left[\frac{1}{2}\partial_{\mu}\varphi\partial^{\mu}\varphi -V(\varphi)\right].
\end{equation}

\end{document}